\documentclass[runningheads]{llncs}
\newcommand{\etal}{et al.}

\usepackage[a4paper,margin=1.2in]{geometry}

\usepackage{silence}
\usepackage[font=small, labelfont=bf]{caption}

\usepackage[T1]{fontenc}
\usepackage[utf8]{inputenc}
\usepackage{acro}
\usepackage{adjustbox}
\usepackage{booktabs}
\usepackage{comment}
\usepackage{todonotes}
\usepackage{colortbl}

\newcolumntype{a}{>{\columncolor{gray!10!white}}c}
\newcolumntype{x}{>{\columncolor{green!10!white}}c}
\newcolumntype{y}{>{\columncolor{blue!10!white}}c}
\newcolumntype{z}{>{\columncolor{yellow!10!white}}c}
\newcolumntype{v}{>{\columncolor{red!10!white}}c}
\definecolor{OliveGreen}{rgb}{0,0.6,0}
\definecolor{ForestGreen}{RGB}{34,139,34}
\definecolor{myblue}{RGB}{37,165,203}
\definecolor{FAUblue}{rgb}{0.000, 0.2196, 0.3961}
\definecolor{myred}{RGB}{175,32,67}

\colorlet{backgroundcol}{cyan!10!white}

\usepackage{graphicx}
\usepackage{enumitem}
\usepackage{color}
\usepackage{xcolor}
\usepackage{listings}
\definecolor{codegreen}{rgb}{0,0.6,0}
\definecolor{codegray}{rgb}{0.5,0.5,0.5}
\definecolor{codepurple}{rgb}{0.58,0.5,0.82}
\definecolor{backcolour}{rgb}{0.95,0.95,0.92}

\lstdefinestyle{mystyle}{
    backgroundcolor=\color{white},   
    commentstyle=\color{codegreen},
    keywordstyle=\color{magenta}\bfseries,
    moredelim=[is][\color{magenta}\bfseries]{@}{@},
    numberstyle=\tiny\color{codegray},
    stringstyle=\color{codepurple},
    basicstyle=\C\footnotesize,
    frame           = tb,         
    framerule       = 0.6pt,      
    rulecolor       = \color{black},
    framesep        = 0.4em,      
    xleftmargin     = 2em,
    framexleftmargin= 2em,
    breakatwhitespace=false,         
    breaklines=true,                 
    captionpos=b,                    
    keepspaces=true,                 
    numbers=left,                    
    numbersep=5pt,                  
    showspaces=false,                
    showstringspaces=false,
    showtabs=false,                  
    tabsize=2
}

\usepackage{amsmath,amssymb,amsfonts}

\usepackage{wrapfig}
\usepackage{graphicx}
\usepackage{soul}

\usepackage{multirow}

\begin{document}

\title{Accelerating Particle-in-Cell Monte Carlo Simulations \\ with MPI, OpenMP/OpenACC and Asynchronous Multi-GPU Programming}

%
\titlerunning{Accelerating Particle-in-Cell Monte Carlo Simulations}
%
\author{Jeremy J. Williams\thanks{Corresponding location: Lindstedtsvägen 5, SE-100 44, Stockholm, Sweden \\ E-mail address: jjwil@kth.se (Jeremy J. Williams)}\inst{1}\and
Felix Liu\inst{1} \and
Jordy Trilaksono\inst{2} \and
David Tskhakaya \inst{3} \and
Stefan Costea\inst{4} \and
Leon Kos \inst{4} \and
Ales Podolnik \inst{3} \and
Jakub Hromadka\inst{3} \and
Pratibha Hegde\inst{1} \and
Marta Garcia-Gasulla \inst{5} \and
Valentin Seitz\inst{5} \and
Frank Jenko \inst{2} \and
Erwin Laure \inst{6} \and
Stefano Markidis\inst{1} }
\authorrunning{Jeremy J. Williams et al.}
%
\institute{KTH Royal Institute of Technology, Stockholm, Sweden \and
Max Planck Institute for Plasma Physics, Garching, Germany \and
Institute of Plasma Physics of the CAS, Prague, Czech Republic  \and 
LECAD Laboratory, University of Ljubljana, Ljubljana, Slovenia \and 
Barcelona Supercomputing Center, Barcelona, Spain \and
Max Planck Computing and Data Facility, Garching and Greifswald, Germany}
\maketitle              
\begin{abstract}
As fusion energy devices advance, plasma simulations play a critical role in fusion reactor design. Particle-in-Cell Monte Carlo simulations are essential for modelling plasma-material interactions and analysing power load distributions on tokamak divertors. Previous work introduced hybrid parallelization in BIT1 using MPI and OpenMP/OpenACC for shared-memory and multicore CPU processing. In this extended work, we integrate MPI with OpenMP and OpenACC, focusing on asynchronous multi-GPU programming with OpenMP Target Tasks using the "nowait" and "depend" clauses, and OpenACC Parallel with the "async(n)" clause. Our results show significant performance improvements: 16 MPI ranks plus OpenMP threads reduced simulation runtime by 53\% on a petascale EuroHPC supercomputer, while the OpenACC multicore implementation achieved a 58\% reduction compared to the MPI-only version. Scaling to 64 MPI ranks, OpenACC outperformed OpenMP, achieving a 24\% improvement in the particle mover function. On the HPE Cray EX supercomputer, OpenMP and OpenACC consistently reduced simulation times, with a 37\% reduction at 100 nodes. Results from MareNostrum 5, a pre-exascale EuroHPC supercomputer, highlight OpenACC's effectiveness, with the "async(n)" configuration delivering notable performance gains. However, OpenMP asynchronous configurations outperform OpenACC at larger node counts, particularly for extreme scaling runs. As BIT1 scales asynchronously to 128 GPUs, OpenMP asynchronous multi-GPU configurations outperformed OpenACC in runtime, demonstrating superior scalability, which continues up to 400 GPUs, further improving runtime. Speedup and parallel efficiency (PE) studies reveal OpenMP asynchronous multi-GPU achieving an 8.77$\times$ speedup (54.81\% PE) and OpenACC achieving an 8.14$\times$ speedup (50.87\% PE) on MareNostrum 5, surpassing the CPU-only version. At higher node counts, PE declined across all implementations due to communication and synchronization costs. However, the asynchronous multi-GPU versions maintained better PE, demonstrating the benefits of asynchronous multi-GPU execution in reducing scalability bottlenecks. While the CPU-only implementation is faster in some cases, OpenMP's asynchronous multi-GPU approach delivers better GPU performance through asynchronous data transfer and task dependencies, ensuring data consistency and avoiding race conditions. Using NVIDIA Nsight tools, we confirmed BIT1's overall efficiency for large-scale plasma simulations, leveraging current and future exascale supercomputing infrastructures. Asynchronous data transfers and dedicated GPU assignments to MPI ranks enhance performance, with OpenMP’s asynchronous multi-GPU implementation utilizing OpenMP Target Tasks with "nowait" and "depend" clauses outperforming other configurations. This makes OpenMP the preferred application programming interface when performance portability, high throughput, and efficient GPU utilization are critical. This enables BIT1 to fully exploit modern supercomputing architectures, advancing fusion energy research. MareNostrum 5 brings us closer to achieving exascale performance.

\end{abstract}


\section{Introduction}
Particle-in-Cell (PIC) Monte Carlo (MC) simulations are vital for understanding complex interactions between plasma and wall materials, which present significant modeling challenges, including the need of resolving different simulation time and spatial scales or modeling accurately atomic and collision processes.

\begin{wrapfigure}{r}{0.4\textwidth}
\centering
\includegraphics[width=\linewidth]{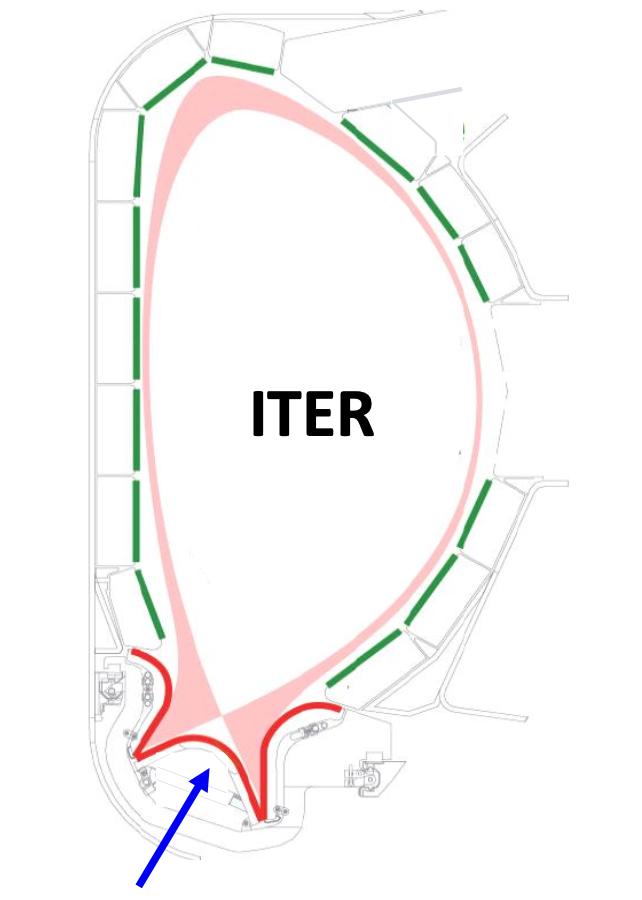}
\caption{BIT1 simulates plasma behavior in the tokamak divertor region (blue arrow), such as in the ITER fusion device.}
\label{divertor}
\end{wrapfigure}

These challenges are particularly notable when modeling plasma-loaded divertors in fusion devices like in the ITER tokamak, a major nuclear fusion project. The divertor manages heat and particle fluxes during tokamak operation. High-energy neutrons produced in a fusion reaction can damage the tokamak’s first wall. Additionally, impurities from the plasma must be efficiently removed to maintain optimal fusion conditions. The divertor diverts plasma flow to a specific region, known as the divertor region, typically located at the bottom of the toroidal chamber (see Fig.~\ref{divertor}).

Among the tools used to address these challenges, BIT1 is a specialized plasma simulation tool, focusing on describing accurately atomic processes and collisions in plasmas during plasma-wall interactions. In particular, BIT1 is widely used to analyze how power is distributed on divertors in these devices. BIT1 plays a critical role as a massively parallel PIC code for studying complex plasma systems and their interactions with various materials. 

Initially introduced by D. Tskhakaya and collaborators~\cite{tskhakaya2007optimization,tskhakaya2010pic}, BIT1 has distinctive capabilities. It models plasmas confined between two conducting walls and includes collision modeling to capture complex plasma dynamics. What makes BIT1 unique is its capability of modeling accurately processes occurring at the interface of plasma and a wall, such as sputtering from the wall, emissions, and collisions. While it has shown that BIT1 is scalable for thousands MPI processes~\cite{williams2023leveraging}, it has two major limitations. The first one is that BIT1 relies only on MPI for parallel communication, even for on-node communication, where a shared-memory computing approach is more convenient and can decrease the memory usage and allow for task-based approaches. The second limitation is the lack of support for running on GPU-accelerated supercomputers. Given the fact that most of the top supercomputers in the world, such as Frontier, Aurora, Eagle and LUMI, the lack of support for GPUs is a major limitation that hinders the usage of BIT1 in the largest supercomputers available. Recent work~\cite{williams2024optimizing} addressed these BIT1 limitations by designing and implementing hybrid parallelization in BIT1 using MPI and OpenMP/OpenACC to enable shared-memory parallelism and multicore CPU processing, alongside the first GPU porting effort aimed at accelerating PIC MC simulations. 

Previously presented findings in~\cite{williams2023leveraging,williams2024understandingchallenges} led to a detailed investigation of BIT1 code performance, identifying performance bottlenecks and outlining a roadmap for optimization~\cite{williams2024optimizing}. This work highlighted areas for enhancing BIT1’s performance, suggesting an initial focus on optimizing the particle mover function, particularly the particle pusher, due to the challenges of arranging particles into cells and MPI ranks. This extended work aims to further enhance BIT1's performance by optimizing its computational processes asynchronously, with a specific emphasis on improving the particle mover function, which is one of the most demanding components of the code. To achieve this, we again employ advanced parallel programming models, OpenMP and OpenACC, to effectively leverage the capabilities of multicore CPUs and heterogeneous hardware, including asynchronous multi-GPU acceleration. By integrating MPI with OpenMP and OpenACC, focusing on asynchronous multi-device programming, we accelerate PIC MC simulations with task-based parallelism (discovered in~\cite{williams2024optimizing}) and ordered execution based on data dependencies while also utilizing multiple execution queues for queue management. These strategies enable concurrent processing, allowing researchers to fully exploit modern computing architectures for large-scale plasma simulations, thereby advancing both the understanding of plasma dynamics and the progress of fusion energy research.

The contributions of this work are the following:
\begin{itemize}[leftmargin=*]
\item We design and implement hybrid MPI+OpenMP and MPI+OpenACC versions of the BIT1 code to improve performance on a single node and in strong scaling tests. This implementation employs a task-based approach to address potential issues with load imbalance in both short and long runs. 
\item We develop and investigate the first GPU porting of the BIT1 code to NVIDIA GPUs using OpenMP and OpenACC in the particle mover stage.
\item We extend our GPU porting efforts to create the first asynchronous multi-GPU implementation, utilizing OpenMP Target Tasks with the "nowait" and "depend" clauses and OpenACC parallel regions using the "async(n)" clause, where n specifies the queue number for concurrent executions.
\item We critically analyze and discuss the performance of the newly ported and further accelerated BIT1 code, demonstrating significant performance improvements, new insights into using asynchronous multi-GPUs, and identifying the next performance optimization steps.
\end{itemize}

The remainder of this paper is organized as follows. Section~2 provides background information on PIC methods, a PIC MC code (BIT1), BIT1's algorithm, previous implementations, and current strategies. Section~3 details our methodology and experimental setup, including modifications made to BIT1 for this extended work. Performance results are presented in Section~4, including hybrid BIT1, porting BIT1 to NVIDIA GPUs for the first time, accelerating BIT1 with asynchronous Multi-GPU programming, and profiling results from NVIDIA Nsight Systems. Related work is discussed in Section~5. Finally, Section~6 offers the discussion, conclusion, and future work.  


\section{Background}
The Particle-in-Cell (PIC) method is among the most commonly used techniques for plasma simulations. They find applications in a diverse range of plasma environments, from space and astrophysical plasma to laboratory settings, industrial processes, and fusion devices. 

\begin{figure*}[!ht]
    \begin{center}
        \includegraphics[width=\textwidth]{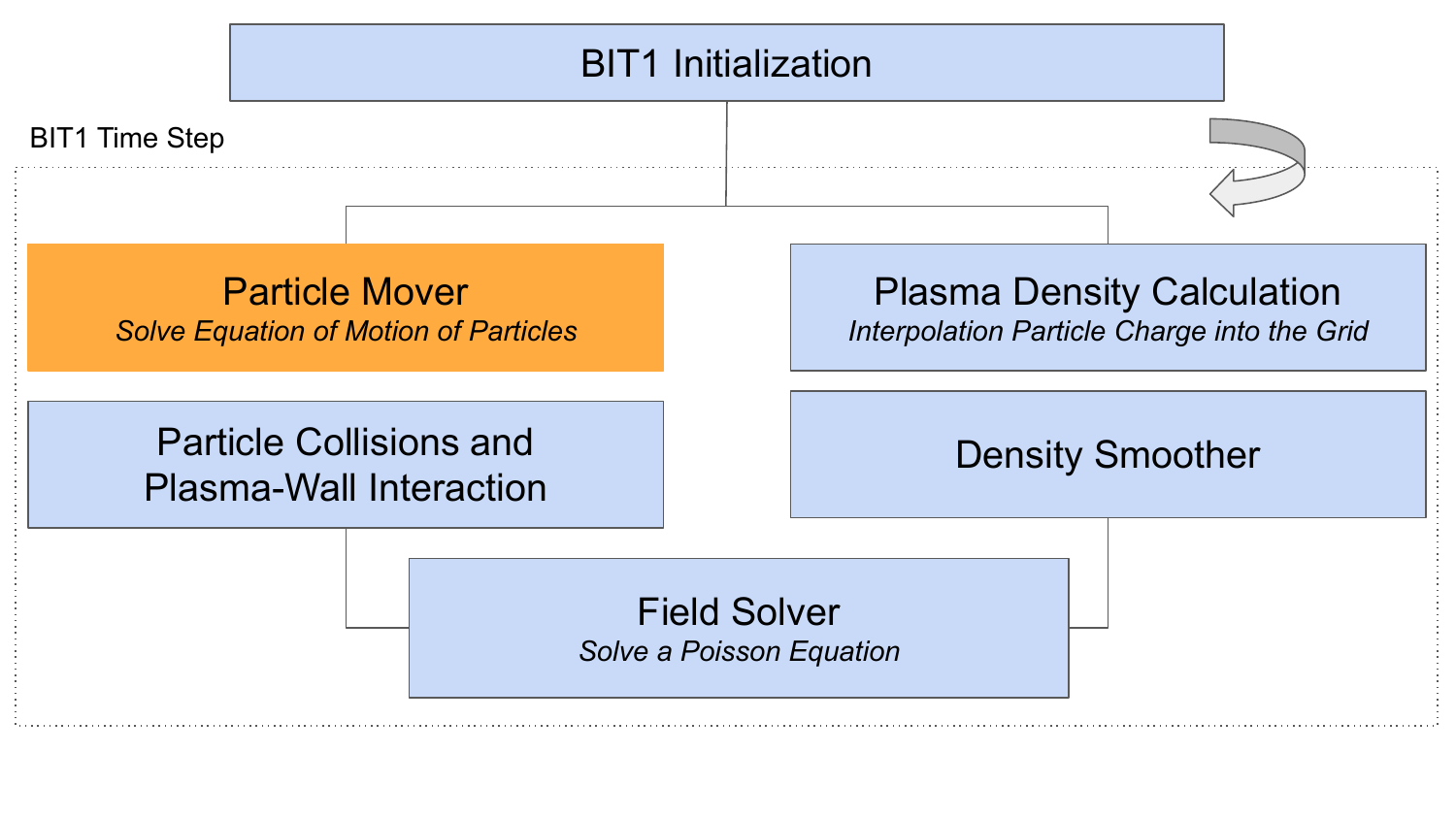}
        \caption{A diagram representing the algorithm used in BIT1. After the initialization the PIC algorithm cycle is repeated at each time step. In orange, we highlight the particle mover step that we parallelize with OpenMP and OpenACC.} 
        \label{diagram}
     \end{center}
\end{figure*}

BIT1 is a 1D3V PIC code, allowing simulations in one dimension while considering particles with three-dimensional velocities. The foundation of BIT1's computational approach is the PIC method, widely adopted in plasma physics. This method involves tracking the trajectories of millions of particles within a field consistent with density and current distributions, while abiding to Maxwell's and Poisson equations. Fig.~\ref{diagram} provides a visual representation of BIT1's explicit PIC method, with its computational cycle consisting of five phases:
\begin{itemize}
    \item \textbf{Plasma Density Calculation}. In this phase, interpolation functions deposit the particle charge on the grid. After this phase, a charge density ($\rho$) is associated with each grid cell. 
    \item \textbf{Density Smoother}. A filtering technique is applied to density on the grid to remove spurious high-frequencies (due to the finite relatively small number of particles to represent the phase space) present in charge density.
    \item \textbf{Field Solver.} At each time step, each cell's electrostatic potential $\Phi$ is calculated by solving the Poisson equation: $\nabla^2\Phi = - \rho/ \epsilon_0$, where $\epsilon_0$ is the vacuum dielectric constant. The Poisson equation is solved using a linear solver (in 1D, the matrix to invert for the solution is a tridiagonal matrix). After the electrostatic potential is calculated, the Electric field is computed as $E = -\nabla \Phi$.
    \item \textbf{Particle Collisions and Plasma-Wall Interaction.} BIT1 also includes advanced models for mimicking different kinds of collisions and the plasma interaction with the wall. A Montecarlo technique is used to model collisions and ionization.
    \item \textbf{Particle Mover.} Each plasma particle (either electrons or ions) is advanced in time, solving first the Newton equation to calculate the $dv_p/dt = q_p/m_p E_p \Delta t$, where $v_p$ is the particle velocity,  $q_p/m_p$ is the charge-to-mass ratio for the particle $p$, and $E_p$ is the electric field at the particle position. After the new velocity is updated, we can update the particle coordinate simply by solving $dx_p/dt = v_p$, where $x_p$ is the particle coordinate.
\end{itemize}

BIT1 is the successor to the \texttt{XPDP1} code, developed by Verboncoeur's team at the University of California, Berkeley (UC Berkeley) in the 1990s~\cite{verboncoeur1993simultaneous}. It comprises approximately 31,600 lines of code and is entirely written in C. Currently, BIT1 does not rely on external numerical libraries but instead implements native solvers for the Poisson equation, particle mover, and smoother.

As shown in Fig.~\ref{diagram}, BIT1 initiates the PIC simulations by configuring the computational grid and setting particle positions and velocities for various species, including electrons and ions. Subsequently, the computational cycle iteratively updates the electric field, particle positions, and velocities, accurately representing the dynamic interactions within the plasma. BIT1 employs advanced Monte Carlo techniques to simulate collisions and ionization processes. Notably, one of the most computationally intensive stages in BIT1 is the particle mover~\cite {williams2023leveraging}, which calculates the trajectories of millions of particles ~\cite{tskhakaya2010pic}.

Unlike many well-established PIC codes designed for high-performance plasma simulations, such as Smilei~\cite{derouillat2018smilei}, an open-source, user-friendly PIC code optimized for supercomputers and applied to a wide range of physics studies, including relativistic laser-plasma interactions and astrophysical plasmas, or Warp-X~\cite{vay2018warp}, a next-generation exascale computing platform for beam-plasma simulations developed as part of the U.S. Department of Energy's Exascale Computing Project, these codes are not optimized for high-density plasma simulations. BIT1, on the other hand, is optimized for plasma edge modeling, specifically the Scrape-Off Layer (SOL), which is the region at the edge of the confined plasma in fusion devices, located just outside the last closed magnetic flux surface. The SOL plays a crucial role in determining how particles and heat are transported to the divertor and plasma-facing components in tokamaks and stellarators.

\begin{figure*}[!ht]
    \begin{center}
        \includegraphics[width=\textwidth]{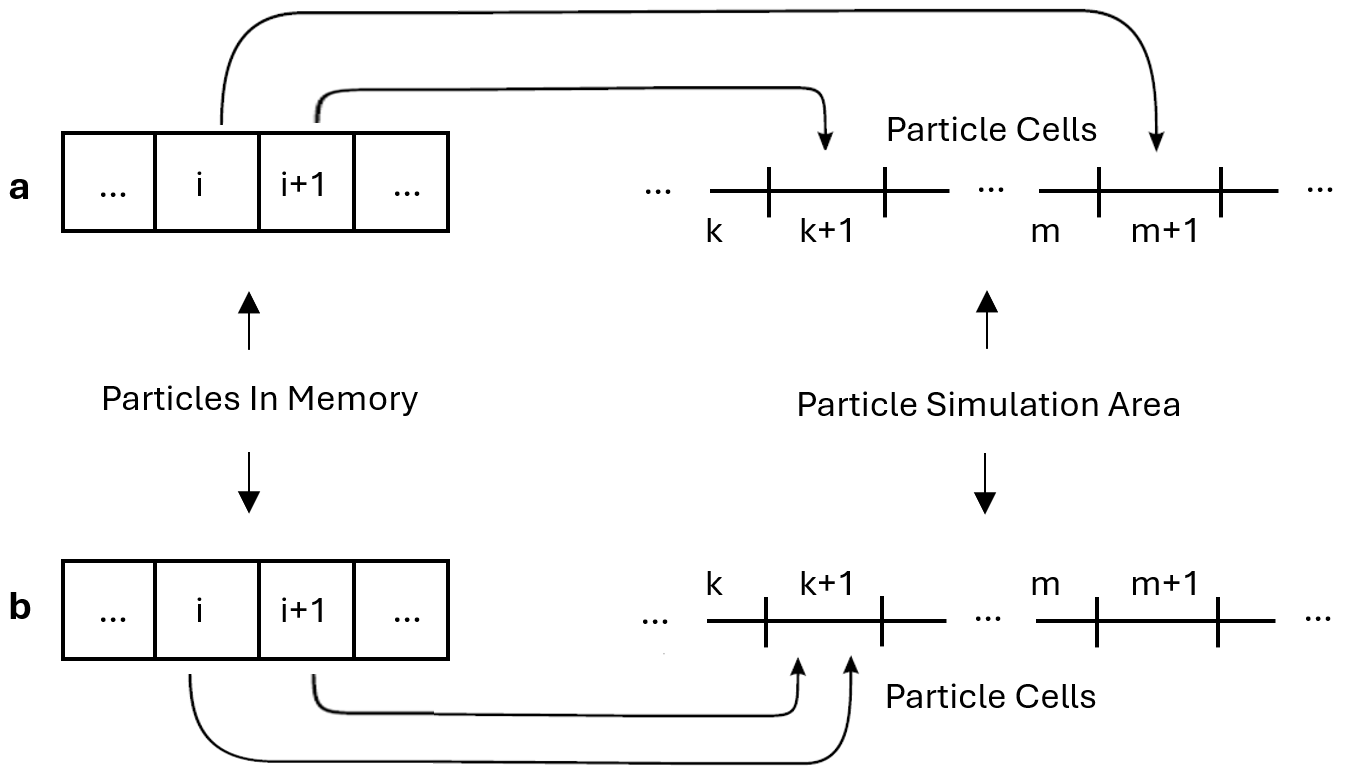}
        \caption{A simple diagram showing two neighboring particles in conventional PIC codes (a) and in BIT1's new approach (b), where in (a), particles that are neighbors in space may not be adjacent in memory, whereas in (b), particles that are neighbors in space are also adjacent in memory and organized according to spatial cells.} 
        \label{BIT1_sorting}
     \end{center}
\end{figure*}

BIT1 is designed to address the unique challenges of SOL modeling, including kinetic effects, plasma-wall interactions, and sheath physics. It incorporates nonlinear plasma, neutral, and impurity interactions through Direct Simulation Monte Carlo (DSMC) collision operators~\cite{tskhakaya2023implementation}. As the first PIC code applied to SOL simulations~\cite{tskhakaya2012recent}, BIT1 efficiently models plasma behavior in this critical edge region. Moreover, BIT1 is tailored for highly collisional plasma modelling and uses novel collision operators, such as the "natural sorting" method~\cite{tskhakaya2007optimization}, which is implemented in BIT1. Its main and distinctive feature is its data layout, where particle information, such as positions and velocities, is stored in memory with a spatial grid cell identifier, as shown in Fig.~\ref{BIT1_sorting}. The particles are associated with the cells they occupy, meaning each grid cell has an associated list of particle information. When a particle moves from one cell to a neighbouring one, its information is removed from one list and added to another. This method can speed up collision operators by a factor of five. However, "natural sorting" requires more memory than the conventional approach, as additional free space must be allocated in the array for each cell. In conventional PIC simulations, extra memory is required only for particles in the entire simulation region, rather than for each individual cell. While this increased memory usage was negligible in the CPU version of BIT1, it can become a bottleneck when offloading to the GPU. Additionally, in plasma simulations, regions of space can experience high particle concentration, leading to situations where some cells contain many particles while others contain only a few. This can cause workload imbalance in the particle mover.

In the current implementation, BIT1 employs domain decomposition for parallelization, utilizing MPI for efficient parallel communication. MPI point-to-point communication is essential for managing information exchange at domain boundaries, crucial for tasks like the smoother, Poisson solver, and handling particles exiting the computational domain. However, the existing BIT1 implementation relies solely on MPI, lacking support for hybrid parallel computing capabilities, such as MPI+OpenMP, MPI+OpenACC, GPU offloading or acceleration. 


\section{Methodology \& Experimental Setup}
In this work, our focus is on accelerating Particle-in-Cell Monte Carlo simulations by porting the BIT1 particle mover to leverage MPI for distributed memory parallelism, OpenMP for shared memory, OpenACC for multicore CPUs, and OpenMP/OpenACC for initial GPU offloading, with further extensions to the GPU offloading implementations using asynchronous multi-GPU programming.

\subsection{Hybrid MPI and OpenMP/OpenACC BIT1} 
 \noindent \textbf{OpenMP Tasks Particle Mover Parallelization.} OpenMP is one of the most widely used programming model designed to facilitate shared-memory parallel programming in high-performance computing (HPC) environments. The OpenMP standard is supported by major compiler suites, including GCC and LLVM, making it accessible to a broad range of developers.

\begin{lstlisting}[
    language=C,
    style=mystyle,
    basicstyle=\fontsize{7}{7}\selectfont\ttfamily,
    caption={Simplified C code snippet illustrating the OpenMP parallelization for CPU in the particle mover.},
    captionpos=b,
    label=1st:openmp_tasks_snippet,
    float=!ht,
    tabsize=2,
    lineskip=0pt,
    xleftmargin=2em, 
    framexleftmargin=2em
]
@#pragma@ omp parallel shared(chsp, sn2d, dinj, nstep, np, x, yp, vx, vy)
    private(isp, i, j) firstprivate(nsp, nc)
{
    @#pragma@ omp single
    {
        for (isp = 0; isp < nsp; isp++) {
            if (chsp[isp]) {
                if (dinj[isp]) {
                    @#pragma@ omp taskloop grainsize(500) nogroup
                        for (j = 0; j < nc; j++) {
                            @#pragma@ omp simd
                                for (i = 0; i < np[isp][j]; i++) {
                                    x[isp][j][i] += nstep[isp] * vx[isp][j][i];
                                }
                        }
                } else {
                    if (dinj[isp]) {
                        if (sn2d[isp]) {
                            @#pragma@ omp taskloop grainsize(500) nogroup
                                for (j = 0; j < nc; j++)
                                    #pragma omp simd
                                        for (i = 0; i < np[isp][j]; i++) {
                                            x[isp][j][i] += nstep[isp] * vx[isp][j][i];
                                            yp[isp][j][i] += nstep[isp] * vy[isp][j][i];
                                        }
                        } else {
                            @#pragma@ omp taskloop grainsize(500) nogroup
                                for (j = 0; j < nc; j++) {
                                    @#pragma@ omp simd
                                        for (i = 0; i < np[isp][j]; i++) {
                                            x[isp][j][i] += nstep[isp] * vx[isp][j][i];
                                        }
                                }
                        }
                    }
                }
            }
        }
    }
}
\end{lstlisting} 

Listing~\ref{1st:openmp_tasks_snippet} showcases our OpenMP port of the core computational function for the particle mover. In the code, \texttt{x[][][]} and \texttt{vx[][][]} represent particle position and velocity in one dimension. \texttt{nsp} is the number of plasma species, and \texttt{nc} is the number of cells in the one-dimensional grid. \texttt{np[][]} denotes the number of particles per species per cell.

Our parallelization approach uses the OpenMP \texttt{taskloop} construct. The outermost loop, with a small number of iterations (species present in the simulation), is deemed unsuitable for traditional parallelization methods. The \texttt{taskloop} construct dynamically distributes loop iterations among available threads, ensuring effective load balancing and optimizing the parallelization strategy for enhanced performance on multicore CPUs.

When examining the code, the \texttt{\#pragma omp parallel} pragma initiates a parallel region with shared variables (\texttt{chsp}, \texttt{dinj}, \texttt{sn2d}, \texttt{nstep}, \texttt{np}, \texttt{x}, \texttt{yp}, \texttt{vx}, \texttt{vy}), while \texttt{isp} and \texttt{i} are private to each thread. The \texttt{firstprivate} clause ensures private and initialized values for \texttt{nsp} and \texttt{nc} for each parallel thread.

Within the parallel region, the \texttt{single} construct ensures that the subsequent block of code is executed by a single thread, crucial for the initialization section. The \texttt{taskloop grainsize(500) nogroup} pragma parallelizes the subsequent loop, dividing iterations into tasks, while optimizing task granularity for efficient parallel execution based on empirical testing and adjustments to achieve optimal performance. The \texttt{nogroup} clause allows for dynamic scheduling.

Finally, the \texttt{simd} pragma within the innermost loop exploits SIMD parallelism, improving vectorization and the efficiency of particle movement calculations.

\noindent \textbf{OpenACC Multicore Particle Mover Parallelization.} OpenACC, akin to OpenMP, is a directive-based programming model primarily designed for GPU accelerators. However, the directive-based approach for GPU offloading can also be advantageous for CPUs, offering a straightforward method for porting codes to CPUs with minimal code changes. 

\begin{lstlisting}[
    language=C,
    style=mystyle,
    basicstyle=\fontsize{7}{7}\selectfont\ttfamily,
    caption={Simplified C code snippet illustrating the OpenACC Multicore CPU parallelization in the particle mover.},
    captionpos=b,
    label=2nd:openacc_multicore_snippet, 
    float=!ht,
    tabsize=2,
    xleftmargin=2em, 
    framexleftmargin=2em 
]
@#pragma@ acc parallel loop present(chsp[:lenA], sn2d[:lenA], dinj[:lenA], 
    nstep[:lenA], np[:lenA][:lenB], x[:lenA][:lenB][:lenC], 
    yp[:lenA][:lenB][:lenC], vx[:lenA][:lenB][:lenC], 
    vy[:lenA][:lenB][:lenC])
{
    for (isp = 0; isp < nsp; isp++) {
        if (chsp[isp]) {
            if (dinj[isp]) {
                @#pragma@ acc loop gang vector
                    for (j = 0; j < nc; j++) {
                        @#pragma@ acc loop vector
                            for (i = 0; i < np[isp][j]; i++) {
                                x[isp][j][i] += nstep[isp] * vx[isp][j][i];
                            }
                    }
            }
        } else {
            if (dinj[isp]) {
                if (sn2d[isp]) {
                    @#pragma@ acc loop gang vector
                        for (j = 0; j < nc; j++) {
                            @#pragma@ acc loop vector
                                for (i = 0; i < np[isp][j]; i++) {
                                    x[isp][j][i] += nstep[isp] * vx[isp][j][i];
                                    yp[isp][j][i] += nstep[isp] * vy[isp][j][i];
                                }
                        }
                } else {
                    @#pragma@ acc loop gang vector
                        for (j = 0; j < nc; j++) {
                            @#pragma@ acc loop vector
                                for (i = 0; i < np[isp][j]; i++) {
                                    x[isp][j][i] += nstep[isp] * vx[isp][j][i];
                                }
                        }
                }
            }
        }
    }
}
\end{lstlisting}

Listing~\ref{2nd:openacc_multicore_snippet} shows our optimized OpenACC parallelization for the particle mover on multicore CPUs. The \texttt{\#pragma acc parallel loop} directive initiates concurrent execution, specifying essential data arrays. Utilizing \texttt{gang} and \texttt{vector} directives enhances parallel processing in a nested loop structure (\texttt{\#pragma acc loop gang vector}) for particle and grid index iterations. Particle positions are updated based on \texttt{nstep} and \texttt{vx}. 

This version strategically uses OpenACC directives (\texttt{\#pragma acc loop gang vector} and \texttt{\#pragma acc loop vector}) to optimize multicore CPUs for the particle mover function, enhancing parallel performance in nested loops.

\subsection{Porting BIT1 w/ OpenMP and OpenACC}

\noindent \textbf{Porting BIT1 with OpenMP Target.} Listing~\ref{3rd:openmp_target_snippet} illustrates our use of the OpenMP target construct to parallelize BIT1's particle mover function for GPU offloading. The code strategically employs OpenMP target directives to optimize particle movement computations by offloading them to GPUs.

\begin{lstlisting}[
    language=C,
    style=mystyle,
    basicstyle=\fontsize{7}{7}\selectfont\ttfamily,
    caption={Simplified C code snippet illustrating the OpenMP (OMP target) parallelization with data clauses and array "shape" for GPU porting in the particle mover.},
    captionpos=b,
    label=3rd:openmp_target_snippet, 
    float=!ht,
    tabsize=2,
    xleftmargin=2em, 
    framexleftmargin=2em 
]
// Enter unstructured data region
@#pragma@ omp target enter data map(to: chsp[:lenA], sn2d[:lenA], dinj[:lenA], 
    nstep[:lenA], np[:lenA][:lenB], x[:lenA][:lenB][:lenC], 
    yp[:lenA][:lenB][:lenC], vx[:lenA][:lenB][:lenC], 
    vy[:lenA][:lenB][:lenC])
{
    for (isp = 0; isp < nsp; isp++) {
        if (chsp[isp]) {
            if (dinj[isp]) {
                @#pragma@ omp target teams distribute parallel for 
                private(i, j) thread_limit(256) num_teams(391)
                    for (j = 0; j < nc; j++)
                        @#pragma@ omp simd
                            for (i = 0; i < np[isp][j]; i++) {
                                x[isp][j][i] += nstep[isp] * vx[isp][j][i];
                        }
            }
        } else {
            if (dinj[isp]) {
                if (sn2d[isp]) {
                    @#pragma@ omp target teams distribute parallel for 
                    private(i, j) thread_limit(256) num_teams(391)
                        for (j = 0; j < nc; j++)
                            @#pragma@ omp simd
                                for (i = 0; i < np[isp][j]; i++) {
                                    x[isp][j][i] += nstep[isp] * vx[isp][j][i];
                                    yp[isp][j][i] += nstep[isp] * vy[isp][j][i];
                                }
                } else {
                    @#pragma@ omp target teams distribute parallel for 
                    private(i, j) thread_limit(256) num_teams(391)
                        for (j = 0; j < nc; j++)
                            @#pragma@ omp simd
                                for (i = 0; i < np[isp][j]; i++) {
                                    x[isp][j][i] += nstep[isp] * vx[isp][j][i];
                            }
                }
            }
        }
    }
// Exit unstructured data region
@#pragma@ omp target exit data map(from: x[:lenA][:lenB][:lenC], yp[:lenA][:lenB][:lenC])
}
\end{lstlisting}

The pragma directive \texttt{\#pragma omp target enter data} initiates the data transfer from the host to the GPU, encompassing crucial arrays such as \texttt{chsp}, \texttt{sn2d}, \texttt{dinj}, and the arrays for particle position and velocity (\texttt{x}, \texttt{yp}, \texttt{vx}, \texttt{vy}).

Within the unstructured data mapping region, the \texttt{\#pragma omp target teams distribute parallel for} directive initiates worksharing across multiple levels of parallelism using combined constructs, thereby enabling the parallel execution of the nested loops for particle movement calculations on the GPU. The \texttt{private(i, j)} clause ensures that the loop variables \texttt{i} and \texttt{j} are private to each thread, preventing race conditions. To fine-tune parallelism, the directives \texttt{thread\_limit(256)} and \texttt{num\_teams(391)} are used to set the maximum number of threads per team and the number of teams, respectively, based on the system specifications and workload requirements of our experimental setup.

In the nested loops, the \texttt{\#pragma omp simd} directive provides a hint to the compiler for potential vectorizations of the inner loops, optimizing SIMD parallelism. The calculations, updating particle positions based on velocities and time steps, are concurrently distributed across GPU threads.

Finally, the \texttt{\#pragma omp target exit data} directive ensures seamless transfer of modified data, specifically particle positions, from the GPU back to the host for efficient GPU offloading and parallelization of particle mover computations.

\noindent \textbf{Porting BIT1 with OpenACC Parallel.} OpenACC, designed for GPU acceleration, simplifies the task of offloading functions to GPUs, offering an accessible solution without complex GPU programming. Supported by platforms like NVIDIA and GCC, OpenACC empowers developers to harness GPU parallel processing efficiently.

\begin{lstlisting}[
    language=C,
    style=mystyle,
    basicstyle=\fontsize{7}{7}\selectfont\ttfamily,
    caption={Simplified C code snippet illustrating the OpenACC (ACC parallel) parallelization with data clauses and array "shape" for GPU porting in the particle mover.},
    captionpos=b,
    label=4th:openacc_parallel_snippet, 
    float=!ht,
    tabsize=2,
    xleftmargin=2em, 
    framexleftmargin=2em 
]
// Enter unstructured data region
@#pragma@ acc enter data copyin(chsp[:lenA], sn2d[:lenA], dinj[:lenA], nstep[:lenA], 
    np[:lenA][:lenB], x[:lenA][:lenB][:lenC], yp[:lenA][:lenB][:lenC], 
    vx[:lenA][:lenB][:lenC], vy[:lenA][:lenB][:lenC])
{
    for (isp = 0; isp < nsp; isp++) {
        if (chsp[isp]) {
            if (dinj[isp]) {
                @#pragma@ acc parallel loop gang worker vector vector_length(128)  
                present(np[:lenA][:lenB], nstep[:lenA], x[:lenA][:lenB][:lenC],  
                vx[:lenA][:lenB][:lenC]) firstprivate(nc, isp) private(i, j) 
                    for (j = 0; j < nc; j++)
                        @#pragma@ acc loop
                            for (i = 0; i < np[isp][j]; i++) {
                                x[isp][j][i] += nstep[isp] * vx[isp][j][i];
                            }
            }
        } else {
            if (dinj[isp]) {
                if (sn2d[isp]) {
                    @#pragma@ acc parallel loop gang worker vector vector_length(128)
                    present(np[:lenA][:lenB], nstep[:lenA], x[:lenA][:lenB][:lenC],
                    yp[:lenA][:lenB][:lenC], vx[:lenA][:lenB][:lenC], 
                    vy[:lenA][:lenB][:lenC]) firstprivate(nc, isp) private(i, j)
                        for (j = 0; j < nc; j++)
                            @#pragma@ acc loop
                                for (i = 0; i < np[isp][j]; i++) {
                                    x[isp][j][i] += nstep[isp] * vx[isp][j][i];
                                    yp[isp][j][i] += nstep[isp] * vy[isp][j][i];
                                }
                } else {
                    @#pragma@ acc parallel loop gang worker vector vector_length(128) 
                    present(np[:lenA][:lenB], nstep[:lenA], x[:lenA][:lenB][:lenC], 
                    vx[:lenA][:lenB][:lenC]) firstprivate(nc, isp) private(i, j)
                            for (j = 0; j < nc; j++)
                                @#pragma@ acc loop
                                    for (i = 0; i < np[isp][j]; i++) {
                                        x[isp][j][i] += nstep[isp] * vx[isp][j][i];
                                    }
                }
            }
        }
    }
// Exit unstructured data region
@#pragma@ acc exit data copyout(x[:lenA][:lenB][:lenC], yp[:lenA][:lenB][:lenC])
}
\end{lstlisting}

In Listing~\ref{4th:openacc_parallel_snippet} , we showcase our OpenACC parallelization of the particle mover function for GPU acceleration. Data movement between CPU and GPU is facilitated using \texttt{\#pragma acc enter data} and \texttt{\#pragma acc exit data} directives with \texttt{copyin} and \texttt{copyout} clauses, managing the transfer of relevant arrays (\texttt{chsp}, \texttt{sn2d}, \texttt{dinj}, \texttt{nstep}, \texttt{np}, \texttt{x}, \texttt{yp}, \texttt{vx}, \texttt{vy}).

In the GPU parallel unstructured data region, the \texttt{\#pragma acc parallel loop} directive is employed to parallelize the outer loop over \texttt{j} for components (\texttt{nc}). The \texttt{present} clause ensures availability of specified arrays, while the \texttt{first private} and \texttt{private} clauses handle variables \texttt{nc}, \texttt{isp}, \texttt{i} and \texttt{j} appropriately.

The loop parallelization strategy uses \texttt{gang}, \texttt{worker}, and \texttt{vector} directives. The \texttt{gang} directive divides loop iterations into gangs, potentially assigned to different cores. Within each gang, \texttt{worker} directive enables concurrent execution, and \texttt{vector} directive subdivides each worker, specifying simultaneous processing with \texttt{vector\_length(128)} determining vector size.

The innermost loop over \texttt{i} is parallelized with \texttt{\#pragma acc loop} directive, optimizing GPU capacity for parallel computations on inner loops while minimizing data transfer overhead.

The \texttt{\#pragma acc exit data copyout} directive ensures copied back modified data to the CPU after GPU computations are complete.

\subsection{Accelerating BIT1 with OpenMP and OpenACC}

\noindent \textbf{Accelerating BIT1 w/ OpenMP Target and Task Dependences.} 
Listing~\ref{5th:openmp_target_tasks_snippet} illustrates our approach to accelerating BIT1's particle mover function for GPU offloading through asynchronous programming with OpenMP. The implementation leverages OpenMP target tasks to enable non-blocking operations, enhancing performance in particle movement computations. 

\begin{lstlisting}[
    language=C,
    style=mystyle,
    basicstyle=\fontsize{6}{6}\selectfont\ttfamily,
    caption={Simplified C code snippet illustrating the asynchronous GPU programming using OpenMP (OMP target tasks) with "nowait" and "depend" clauses to further enhance the particle mover.},
    captionpos=b,
    label=5th:openmp_target_tasks_snippet, 
    float=!ht,
    tabsize=2,
    xleftmargin=2em, 
    framexleftmargin=2em 
]
// Enter unstructured data region asynchronously
@#pragma@ omp target enter data nowait map(to: chsp[:lenA], sn2d[:lenA], dinj[:lenA], 
    nstep[:lenA], np[:lenA][:lenB], x[:lenA][:lenB][:lenC], yp[:lenA][:lenB][:lenC], 
    vx[:lenA][:lenB][:lenC], vy[:lenA][:lenB][:lenC]) depend(out: x[:lenA][:lenB][:lenC], 
    yp[:lenA][:lenB][:lenC], nstep[:lenA], np[:lenA][:lenB], vx[:lenA][:lenB][:lenC],
    vy[:lenA][:lenB][:lenC])
{
    for (isp = 0; isp < nsp; isp++) {
        if (chsp[isp]) {
            if (dinj[isp]) {
                @#pragma@ omp target teams distribute parallel for
                    depend(inout: x[:lenA][:lenB][:lenC])
                    depend(in: nstep[:lenA] np[:lenA][:lenB], vx[:lenA][:lenB][:lenC])
                    private(i, j) thread_limit(256) num_teams(391) nowait
                    for (j = 0; j < nc; j++) {
                        @#pragma@ omp simd
                            for (i = 0; i < np[isp][j]; i++) {
                                x[isp][j][i] += nstep[isp] * vx[isp][j][i];
                            }
                    }
            }
        } else {
            if (dinj[isp]) {
                if (sn2d[isp]) {
                    @#pragma@ omp target teams distribute parallel for
                        depend(inout: x[:lenA][:lenB][:lenC], yp[:lenA][:lenB][:lenC])
                        depend(in: nstep[:lenA], np[:lenA][:lenB], vx[:lenA][:lenB][:lenC])
                        private(i, j) thread_limit(256) num_teams(391) nowait
                        for (j = 0; j < nc; j++) {
                            @#pragma@ omp simd
                                for (i = 0; i < np[isp][j]; i++) {
                                    x[isp][j][i] += nstep[isp] * vx[isp][j][i];
                                    yp[isp][j][i] += nstep[isp] * vy[isp][j][i];
                                }
                        }
                } else {
                    @#pragma@ omp target teams distribute parallel for
                        depend(inout: x[:lenA][:lenB][:lenC])
                        depend(in: nstep[:lenA], np[:lenA][:lenB], vx[:lenA][:lenB][:lenC])
                        private(i, j) thread_limit(256) num_teams(391) nowait
                        for (j = 0; j < nc; j++) {
                            @#pragma@ omp simd
                                for (i = 0; i < np[isp][j]; i++) {
                                    x[isp][j][i] += nstep[isp] * vx[isp][j][i];
                                }
                        }
                }
            }
        }
    } 
// Exit unstructured data region asynchronously
@#pragma@ omp target exit data nowait map(from: x[:lenA][:lenB][:lenC], yp[:lenA][:lenB][:lenC])
    depend(in: x[:lenA][:lenB][:lenC], yp[:lenA][:lenB][:lenC])
}
// Wait for all asynchronous tasks and data transfer to complete
@#pragma@ omp taskwait
\end{lstlisting}

The \texttt{\#pragma omp target enter data nowait} directive initiates asynchronous data transfer from the host to the GPU, preventing execution blocking and allowing overlapping computation and communication. This directive maps critical arrays such as \texttt{chsp}, \texttt{sn2d}, \texttt{dinj}, and the arrays for particle positions and velocities (\texttt{x}, \texttt{yp}, \texttt{vx}, \texttt{vy}). The \texttt{depend(out:)} clause in the directive specifies data dependencies, ensuring the correct order of operations across tasks while maintaining asynchronous execution. These dependencies ensure that the updated data (such as particle positions and velocities) is available after the computation on the GPU is completed.

Within the data region, the \texttt{\#pragma omp target teams distribute parallel for} directive parallelizes the loops across GPU teams and threads, enhancing computational efficiency. The \texttt{depend(inout:)} clause ensures task dependencies for synchronization by specifying that the task both reads and writes the data, maintaining consistency and preventing race conditions, while the \texttt{depend(in:)} clause ensures necessary input data is available before GPU computation begins without unnecessary serialization. The \texttt{nowait} clause enables subsequent tasks to execute without waiting for the current task to complete, maximizing concurrency. Including the \texttt{private(i, j)} clause in the \texttt{\#pragma omp target teams distribute parallel for} directive ensures that loop variables are private to each thread, avoiding race conditions. Additionally, the \texttt{thread\_limit(256)} and \texttt{num\_teams(391)} clauses optimize the GPU kernel launch configuration, enabling efficient scaling for the specific problem size and target execution environment.

In the nested loops, the \texttt{\#pragma omp simd} directive provides a hint to the compiler for potential vectorization of the inner loops, optimizing SIMD parallelism. The calculations, updating particle positions based on velocities and time steps, are concurrently distributed across GPU threads.

Finally, the \texttt{\#pragma omp target exit data nowait} directive is used to asynchronously transfer the updated particle positions and velocities back to the host. The \texttt{depend(in:)} clause ensures that the data is correctly synchronized when transferring the results back. This is followed by the \texttt{\#pragma omp taskwait} directive, ensuring the completion of all asynchronous tasks and data transfers before continuing to the next phase of computation.

\vspace{2mm}
\noindent \textbf{Accelerating BIT1 w/ OpenACC Parallel and Queue Management.}  Listing~\ref{6th:openacc_parallel_async_snippet} demonstrates the use of OpenACC for asynchronous GPU programming, enhancing the efficiency of the particle mover function by offloading computations to the GPU while ensuring that data transfers and computations are managed without blocking the CPU.

\begin{lstlisting}[
    language=C,
    style=mystyle,
    basicstyle=\fontsize{6}{6}\selectfont\ttfamily,
    caption={Simplified C code snippet illustrating asynchronous OpenACC parallelization for GPU-accelerated particle mover function with efficient data handling.},
    captionpos=b,
    label=6th:openacc_parallel_async_snippet, 
    float=!ht,
    tabsize=2,
    xleftmargin=2em, 
    framexleftmargin=2em 
]
// Enter unstructured data region asynchronously
@#pragma@ acc enter data copyin(chsp[:lenA], sn2d[:lenA], dinj[:lenA], nstep[:lenA], 
                             np[:lenA][:lenB], x[:lenA][:lenB][:lenC], 
                             yp[:lenA][:lenB][:lenC], vx[:lenA][:lenB][:lenC], 
                             vy[:lenA][:lenB][:lenC]) device_type(nvidia) async(1)
{
    for (isp = 0; isp < nsp; isp++) {
        if (chsp[isp]) {
            if (dinj[isp]) {
                @#pragma@ acc parallel loop gang worker num_workers(32) vector vector_length(32)
                                     present(np[:lenA][:lenB], nstep[:lenA],
                                     x[:lenA][:lenB][:lenC], vx[:lenA][:lenB][:lenC])
                                     firstprivate(nc, isp) private(i, j)
                                     device_type(nvidia) async(2)
                    for (j = 0; j < nc; j++) {
                        @#pragma@ acc loop
                            for (i = 0; i < np[isp][j]; i++) {
                                x[isp][j][i] += nstep[isp] * vx[isp][j][i];
                            }
                    }
            }
        } else {
            if (dinj[isp]) {
                if (sn2d[isp]) {
                    @#pragma@ acc parallel loop gang worker num_workers(32) vector vector_length(32)
                                         present(np[:lenA][:lenB], nstep[:lenA],
                                         x[:lenA][:lenB][:lenC], yp[:lenA][:lenB][:lenC],
                                         vx[:lenA][:lenB][:lenC], vy[:lenA][:lenB][:lenC])
                                         firstprivate(nc, isp) private(i, j)
                                         device_type(nvidia) async(3)
                        for (j = 0; j < nc; j++) {
                            @#pragma@ acc loop
                                for (i = 0; i < np[isp][j]; i++) {
                                    x[isp][j][i] += nstep[isp] * vx[isp][j][i];
                                    yp[isp][j][i] += nstep[isp] * vy[isp][j][i];
                                }
                        }
                } else {
                    @#pragma@ acc parallel loop gang worker num_workers(32) vector vector_length(32)
                                         present(np[:lenA][:lenB], nstep[:lenA], \
                                         x[:lenA][:lenB][:lenC], vx[:lenA][:lenB][:lenC])
                                         firstprivate(nc, isp) private(i, j)
                                         device_type(nvidia) async(4)
                        for (j = 0; j < nc; j++) {
                            @#pragma@ acc loop
                                for (i = 0; i < np[isp][j]; i++) {
                                    x[isp][j][i] += nstep[isp] * vx[isp][j][i];
                                }
                        }
                }
            }
        }
    }
    // Exit unstructured data region asynchronously
    @#pragma@ acc exit data copyout(x[:lenA][:lenB][:lenC], yp[:lenA][:lenB][:lenC])
                           device_type(nvidia) async(1)   
}
// Wait for all asynchronous tasks and data transfer to complete
@#pragma@ acc wait(1, 2, 3, 4)
\end{lstlisting}

In Listing~\ref{6th:openacc_parallel_async_snippet}, the code snippet highlights the asynchronous parallelization of the particle mover function using OpenACC for further GPU acceleration. Data transfer is managed with the \texttt{\#pragma acc enter data} and \texttt{\#pragma acc exit data} directives, which utilize the \texttt{copyin} and \texttt{copyout} clauses to handle the transfer of essential arrays (\texttt{chsp}, \texttt{sn2d}, \texttt{dinj}, \texttt{nstep}, \texttt{np}, \texttt{x}, \texttt{yp}, \texttt{vx}, \texttt{vy}) between the CPU and GPU.

The asynchronous data entry is initiated with \texttt{async(1)}, allowing for non-blocking data transfer while the GPU executes computations. The \texttt{async(n)} clause indicates that the operation should be executed asynchronously using a specific queue identified by the integer \texttt{n}. By specifying \texttt{async(1)}, the data transfer operations are placed in \texttt{queue 1}, enabling the GPU to proceed with its tasks without waiting for these data transfers to complete. Using \texttt{async(1)} in both \texttt{\#pragma acc enter data} and \texttt{\#pragma acc exit data} means that both \texttt{data entry} and \texttt{exit} operations are handled through the same asynchronous queue, ensuring that data transfers are synchronized and managed together. This structure maintains consistency in data handling, ensuring that the GPU has the required data available before computation and that the modified data is returned to the CPU without unnecessary delays.

Within the function, each particle (\texttt{isp}) is processed, and computations are conditional based on the state of \texttt{chsp} and \texttt{dinj}. When particles are moving, the parallelization leverages the \texttt{\#pragma acc parallel loop} directive to handle the outer loop iterating over \texttt{j} for the number of components (\texttt{nc}). The \texttt{present} clause ensures that the specified arrays are available on the GPU during execution, while \texttt{firstprivate} and \texttt{private} clauses manage the variables \texttt{nc}, \texttt{isp}, \texttt{i}, and \texttt{j}. The \texttt{device\_type(nvidia)} directive specifies that the computations will utilize NVIDIA GPUs.

The use of \texttt{gang}, \texttt{worker}, and \texttt{vector} directives in the \texttt{\#pragma acc parallel loop} allows for an efficient division of workload across available GPU cores. Notably, a smaller configuration with \texttt{num\_workers(32)} and \texttt{vector\_length(32)} is employed instead of the previous \texttt{num\_workers(128)}. This choice facilitates better load balancing and avoids oversubscription of GPU resources, which can occur when too many threads are assigned to a single GPU. By using fewer workers and smaller vectors, the program enhances context switching efficiency and maintains optimal performance across varied workloads.

After the computational loops, the \texttt{\#pragma acc exit data copyout} directive ensures that the modified data is sent back to the CPU, marking the completion of GPU computations. The asynchronous \texttt{async(1)} clause allows this operation to occur without blocking other operations.

Finally, the \texttt{\#pragma acc wait(1, 2, 3, 4)} directive synchronizes all asynchronous tasks, ensuring that all computations are complete before the function exits. This efficient use of asynchronous programming with OpenACC, leveraging queue management for the asynchronous calls, allows for improved performance in the particle mover function, reducing data transfer overhead and optimizing GPU utilization. 
%


\subsection{Experimental Setup} 

The following four systems are used in this extended work: 
\begin{itemize}[leftmargin=*]
\item \textbf{Dardel}, an HPE Cray EX supercomputer, features a CPU partition with 1,278 compute nodes. Each node is equipped with two AMD EPYC™ Zen2 2.25 GHz 64-core processors, 256 GB DRAM (58\% of nodes), 512 GB DDR5 (40\% of nodes), 1 TB memory (0.1\% of nodes) and 2 TB memory (1.9\% of nodes), and interconnected using an HPE Slingshot network with Dragonfly topology providing 200 GB/s bandwidth. The Lustre file system has a 12 PB capacity, and the operating system is SUSE Linux Enterprise Server 15 SP3. 
We load GNU compiler suite option “PrgEnv-gnu” for compiler "gcc v11.2.0" and MPI library, “cray-mpich v8.1.17".
\item \textbf{NJ}, an HPC system, has an AMD EPYC 7302P 16-Core Processor with 32 CPU cores. It operates on the x86\_64 platform, 2 threads per core, and 16 cores per socket, running at a 3.0 GHz base clock. \textbf{NJ} also hosts two NVIDIA A100 GPUs with 40 GB HBM2e memory, 6912 Shading Units, 432 Tensor Cores, and 108 SM Count. GPUs have 192 KB L1 Cache per SM, 40 MB L2 Cache, and offer double-precision matrix (FP64) performance of approximately 9.746 TFLOPs. We load CUDA Driver v11.0, NVIDIA HPC SDK v23.7 with GCC v12.2.0 and OpenMPI v4.1.5 for intra-node communication.
\item \textbf{Vega}, a petascale EuroHPC supercomputer, has a CPU partition with 960 compute nodes, each equipped with two AMD EPYC 7H12 64-core processors, 256 GB DDR4 SDRAM (80\% of nodes), and 1 TB DDR4 SDRAM (20\% of nodes). The nodes are interconnected using Mellanox ConnectX-6 InfiniBand HDR100 with a Dragonfly+ topology, providing up to 500 GiB/s bandwidth. \textbf{Vega}'s storage includes a Ceph File System (CephFS) with 23 PB and a Lustre File System (LFS) with 1 PB and 80 OSTs. In addition to the CPU partition, \textbf{Vega} features a GPU partition with 60 GPU nodes, each equipped with 512 GB DDR4, two AMD EPYC 7H12 64-core processors, and four Nvidia A100 GPUs (40 GB HBM2, 6,912 FP32 CUDA cores, 432 Tensor cores per GPU), delivering a peak performance of up to 10.05 PFLOPs. The OS is Red Hat Enterprise Linux 8, and applications are loaded with CUDA Driver v11.7.0, NVIDIA HPC SDK v22.7 with GCC v12.3.0 and OpenMPI v4.1.4 for intra-node communication.
\item \textbf{MareNostrum 5 (MN5)}, a pre-exascale EuroHPC supercomputer, has a GPP (General Purpose Partition) with 6,480 compute nodes. Each node features two Intel Sapphire Rapids 8480+ 56-core processors (72 nodes with Intel Sapphire Rapids HBM), 128 GB HBM (1\% of nodes), 256 GB DDR5 (95\% of nodes), and 1 TB memory (4\% of nodes), interconnected by ConnectX-7 NDR200 InfiniBand (shared by two nodes, 100 Gb/s per node). \textbf{MN5} provides 248 PB of storage with SSD/Flash and hard disks, achieving 1.2 TB/s write and 1.6 TB/s read speeds. It also offers 402 PB of long-term archive storage based on tape. In addition to the GPP, \textbf{MN5} includes 1,120 GPU nodes, each with 512 GB DDR5, two Intel Sapphire Rapids 8460Y+ 40-core processors, and four Nvidia Hopper H100 GPUs (64 GB HBM2 each), delivering up to 295.81 PFLOPs. The OS is Red Hat Enterprise Linux 9.2 (Plow), and applications are loaded with CUDA Driver v12.1.1, NVIDIA HPC SDK v23.7 with GCC v12.3.0 and OpenMPI v4.1.5 for intra-node communication.
\end{itemize}

\begin{table}[h!]
\centering
\resizebox{0.8\textwidth}{!}{ 
    \begin{tabular}{|c|l|}
    \hline
    \multirow{4}{*}{\centering \textbf{BIT1 Test Case}} & Initial uniformly distributed hot plasma \\
    & interacts with initial uniformly distributed \\
    & monoatomic gas, ultimately ionizing the gas \\
    & and cooling down the plasma. \\
    \hline
    \multirow{4}{*}{\centering \textbf{Initial Conditions}} & Three particle species: electrons, D\(^+\) ions, \\
     & and D neutrals are uniformly distributed \\
     & inside the system. Electron and ion initial \\
     & densities: \(10^{21}\, \text{m}^{-3}\) with initial \\
     & temperatures of 20 eV for both charged species. \\
     & Neutral initial density: \(10^{21}\, \text{m}^{-3}\) \\
     & with initial temperature of 1 eV.  \\
     & No external electric or magnetic fields. \\
    \hline
    \multirow{4}{*}{\centering \textbf{Boundary Conditions}} & Periodic boundaries: particles crossing a \\
    & boundary are reinjected from the opposite   \\
    & one with the same velocities, resulting in   \\
    & no plasma-wall interaction. \\
    \hline
    \multirow{2}{*}{\centering \textbf{Computational Mesh}} & 100,000 cells with an initial 100 particles per \\
     & cell per species. System size: 1 m, resulting \\
     & in a cell size of \(10\, \mu\text{m}\) (micrometers).\\
    \hline
    \multirow{3}{*}{\centering \textbf{Simulation Details}} & Simulation ran for 200,000 time steps, with a \\
     & time step of \(4 \times 10^{-14}\, \text{s}\), resulting in a \\
     & total of 8 ns of simulated plasma. Particle \\
     & collisions considered in the simulation \\
     & include electron-neutral elastic, excitation,  \\
     & and ionization processes. \\
    \hline
    \end{tabular}
}
\caption{BIT1 PIC MC Test Case}
\label{table:1}
\end{table}

In this extended work, we focus on optimizing the particle mover function in BIT1, with a particular emphasis on closely monitoring and analyzing BIT1 performance. As shown in Table~\ref{table:1}, we use a test case simulating neutral particle ionization resulting from interactions with electrons in upcoming magnetic confinement fusion devices like ITER and DEMO.  The scenario involves an unbounded unmagnetized plasma consisting of electrons, $D^+$ ions and $D$ neutrals. Due to ionization, neutral concentration decreases with time according to $\partial n / \partial t = n n_e R$, where $n$, $n_e$ and $R$ are neutral particles, plasma densities and ionization rate coefficient, respectively. We use a one-dimensional geometry with 100K cells, three plasma species ($e$ electrons, $D^+$ ions and $D$ neutrals), and 10M particles per species. The total number of particles in the system is 30M. Unless differently specified, we simulate up to 200K time steps. An important point of this test is that it does not use the field solver and smoother phases, as depicted in Fig.~\ref{diagram}.


\section{Performance Results}

\subsection{Hybrid MPI and OpenMP/OpenACC BIT1}
Focusing on both intra-node and inter-node testing, an in-depth investigation into how BIT1 performs in terms of "execution time" has been conducted, to explore the advantages of utilizing our hybrid approaches. The aim was to determine if using both MPI and OpenMP/OpenACC, instead of just MPI, would make a significant difference. 

\begin{figure*}[h!]
    \begin{center}
        \includegraphics[width=0.9\textwidth]{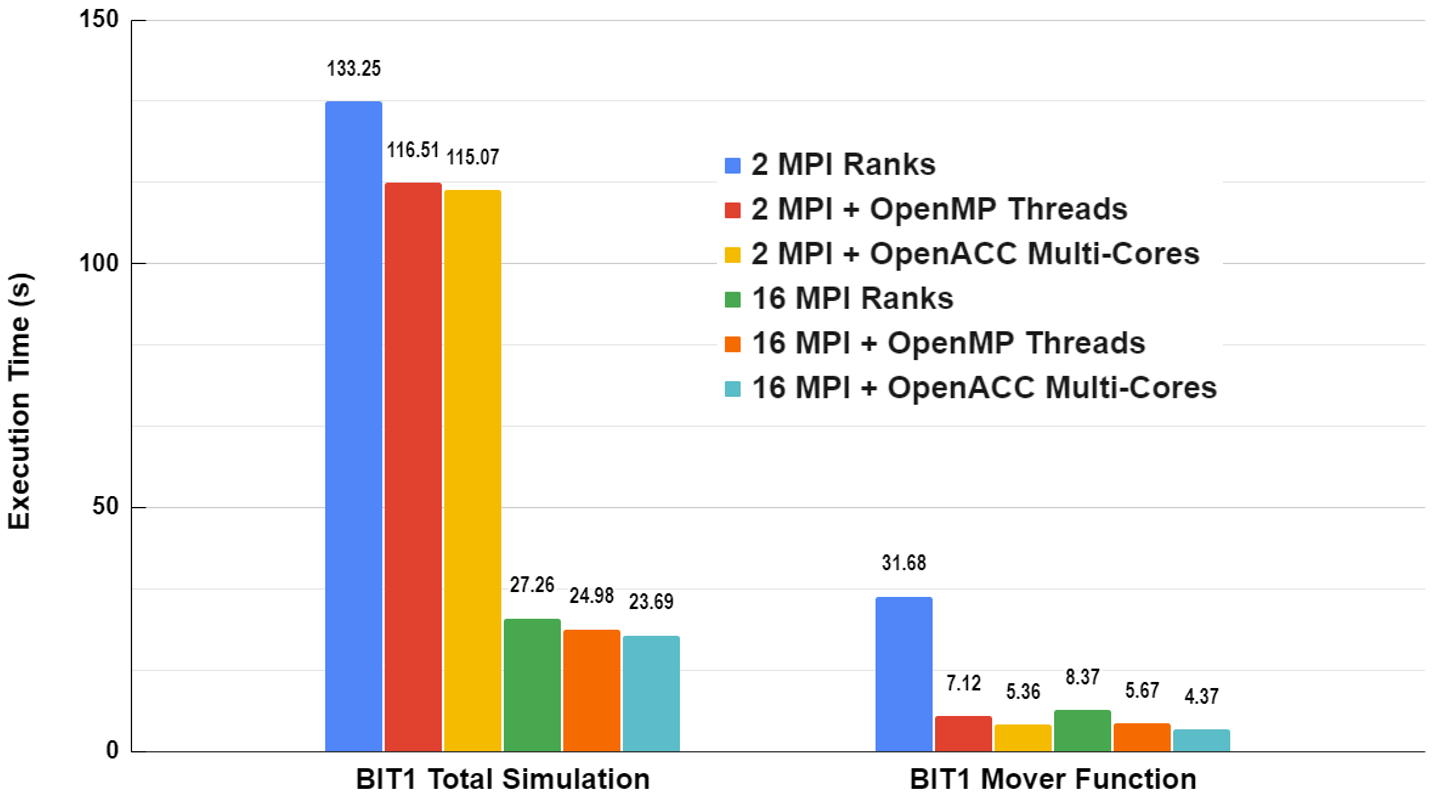}
        \caption{Hybrid BIT1 total simulation and optimized mover function using 2 and 16 ranks per node on \emph{NJ} for 1000 times steps. } \label{BIT_Total_2_and_16_Ranks}
     \end{center}
\end{figure*}
\begin{figure*}[h!]
    \begin{center}
        \includegraphics[width=0.9\textwidth]{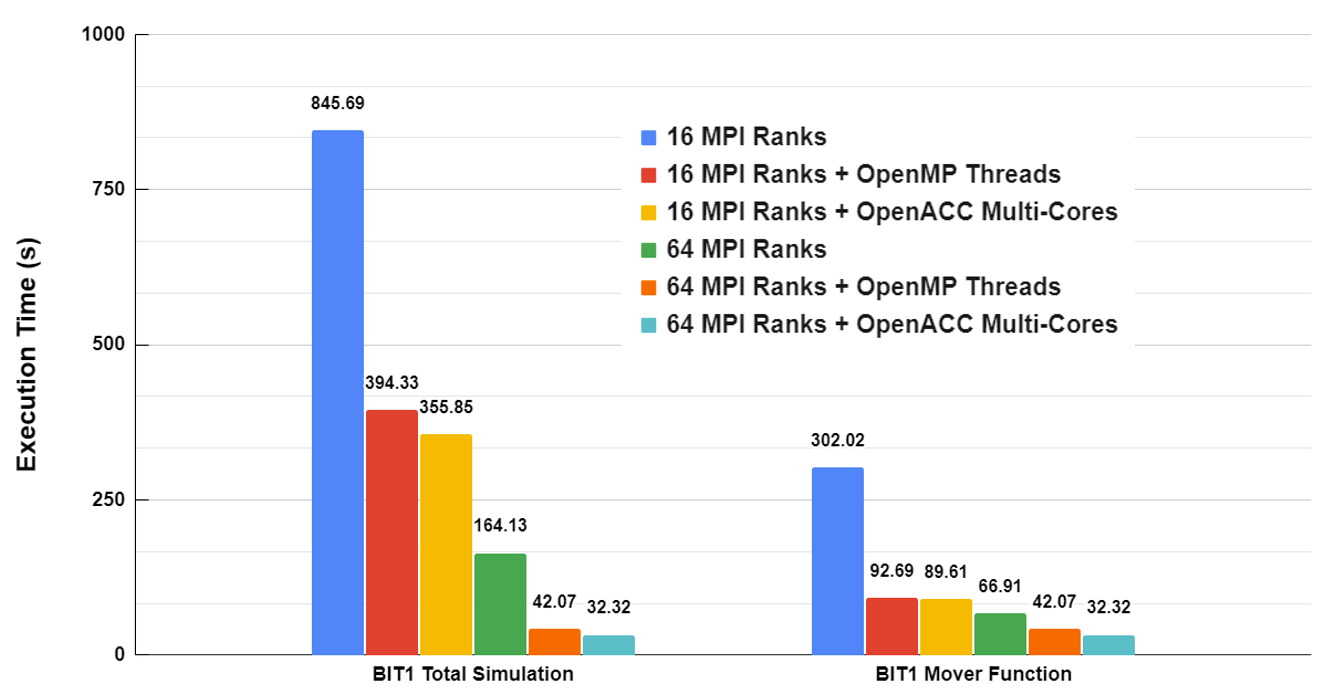}
        \caption{Hybrid BIT1 total simulation and optimized mover function using 16 and 64 ranks per node on \emph{Vega} for 20000 times steps.} \label{BIT_Total_16_and_64_Ranks}
     \end{center}
\end{figure*}

Fig.~\ref{BIT_Total_2_and_16_Ranks} shows executions of hybrid BIT1 total execution vs. optimized mover function using 2 and 16 ranks per node for 1000 times steps on \emph{NJ}. For both 2 ranks and 16 ranks, our hybrid MPI+OpenMP version of BIT1 shows a reduction in both total simulation and mover function time. This suggests that parallelizing BIT1 with OpenMP threads improves performance by enabling multiple threads to work on the problem concurrently. Similar to OpenMP, our hybrid MPI+OpenACC version for multicore CPUs also results in a reduction in both total simulation and mover function time. Extending further on \emph{VEGA} for the purpose of long runs and more time-steps,  Fig.~\ref{BIT_Total_16_and_64_Ranks} illustrates hybrid BIT1 executions for total execution and optimized mover function using 16 and 64 ranks per node for 20,000 time steps. 
Both the hybrid MPI+OpenMP version shows significant reductions in total simulation and mover function times. For example, with 16 MPI ranks, total simulation time decreases from 845.69 seconds to 394.33 seconds, achieving a 53.4\% speedup, while mover function time drops from 302.02 seconds to 92.69 seconds, resulting in a 69.3\% speedup with OpenMP threads. The MPI+OpenACC version on multicore CPUs further reduces total simulation time to 355.85 seconds (57.0\% reduction) and mover function time to 89.61 seconds (70.3\% improvement). Scaling to 64 MPI ranks, total simulation time is reduced to 164.13 seconds and mover function time to 66.91 seconds. Adding OpenMP threads with 64 MPI ranks cuts total simulation and mover function times to 42.07 seconds, yielding a 74.4\% reduction in total simulation time and a 37.2\% reduction in mover function time. Lastly, the hybrid MPI+OpenACC version with 64 MPI ranks achieves the best performance, reducing both times to 32.32 seconds, resulting in an 80.3\% reduction in total simulation time and a 51.7\% reduction in mover function time compared to the 64 MPI ranks baseline. Both Runs on \emph{NJ} and \emph{Vega} demonstrates that BIT1 benefits when used with multicore CPUs, due to better utilization of CPU cores through parallelization.

Investigating the scalability of hybrid BIT1 on CPUs, Fig.~\ref{BIT_Total_2_and_16_Ranks} and~\ref{BIT_Total_16_and_64_Ranks}, it is easy to see that with an increase in MPI ranks from 2 to 64, there is a significant improvement in performance for both total simulation and optimized mover function execution times. This shows that our hybrid BIT1 scales well. However, to confirm these findings, our second system, \emph{Dardel}, was used for further investigation with a focus on the optimized mover function using OpenMP (since OpenACC was not used). 

\begin{figure*}[h!]
    \begin{center}
        \includegraphics[width=0.9\textwidth]{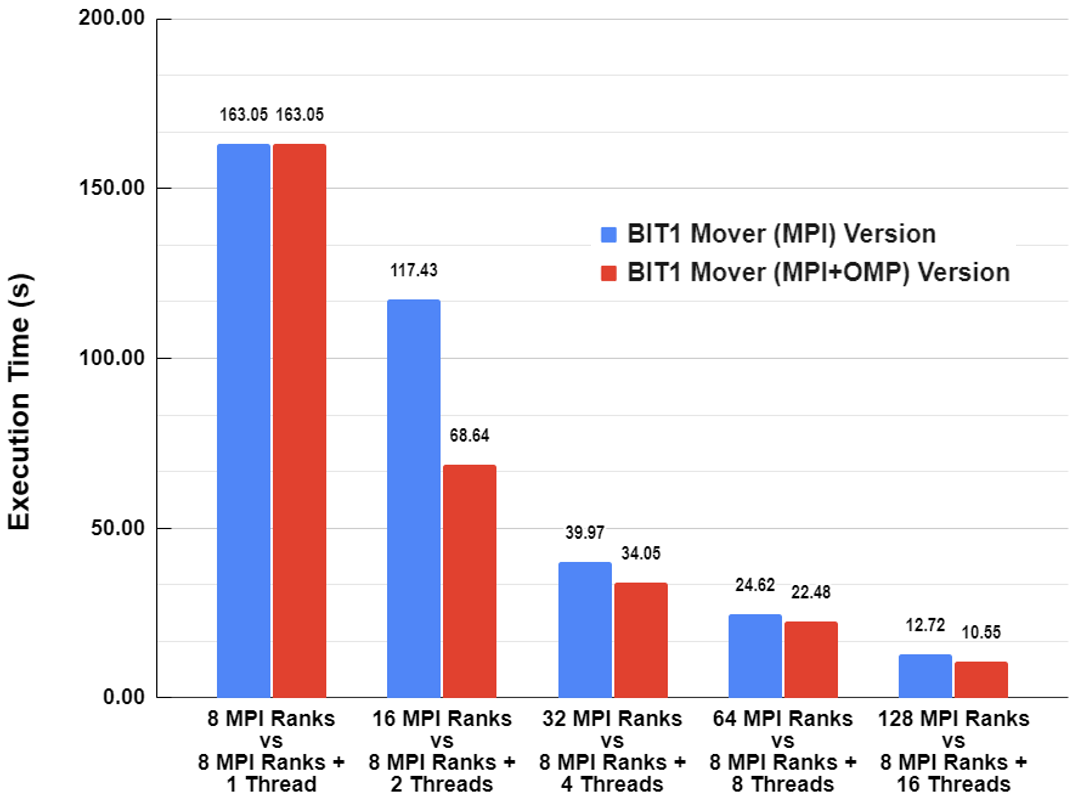}
        \caption{BIT1 Optimized mover function - strong scaling up to 128 MPI ranks on \emph{Dardel} for 1000 times steps.} \label{BIT_Scaling_Dardel}
     \end{center}
\end{figure*}

On \emph{Dardel}, as seen in Fig.~\ref{BIT_Scaling_Dardel}, the execution time significantly decreases as the number of MPI ranks and OpenMP threads increases, showcasing the potential for efficient parallel executions. For instance, with 128 MPI ranks versus 8 MPI ranks with 16 OpenMP threads, the execution time reduces to 10.55 seconds, a notable improvement from the original 12.72 seconds.

Scaling up to 100 nodes and 200,000 time steps, on \emph{Dardel}, our analysis shows notable performance improvements when using hybrid MPI+OpenMP versions of BIT1 for both mover function and total simulation.
\begin{figure*}[h!]
    \begin{center}
        \includegraphics[width=0.9\textwidth]{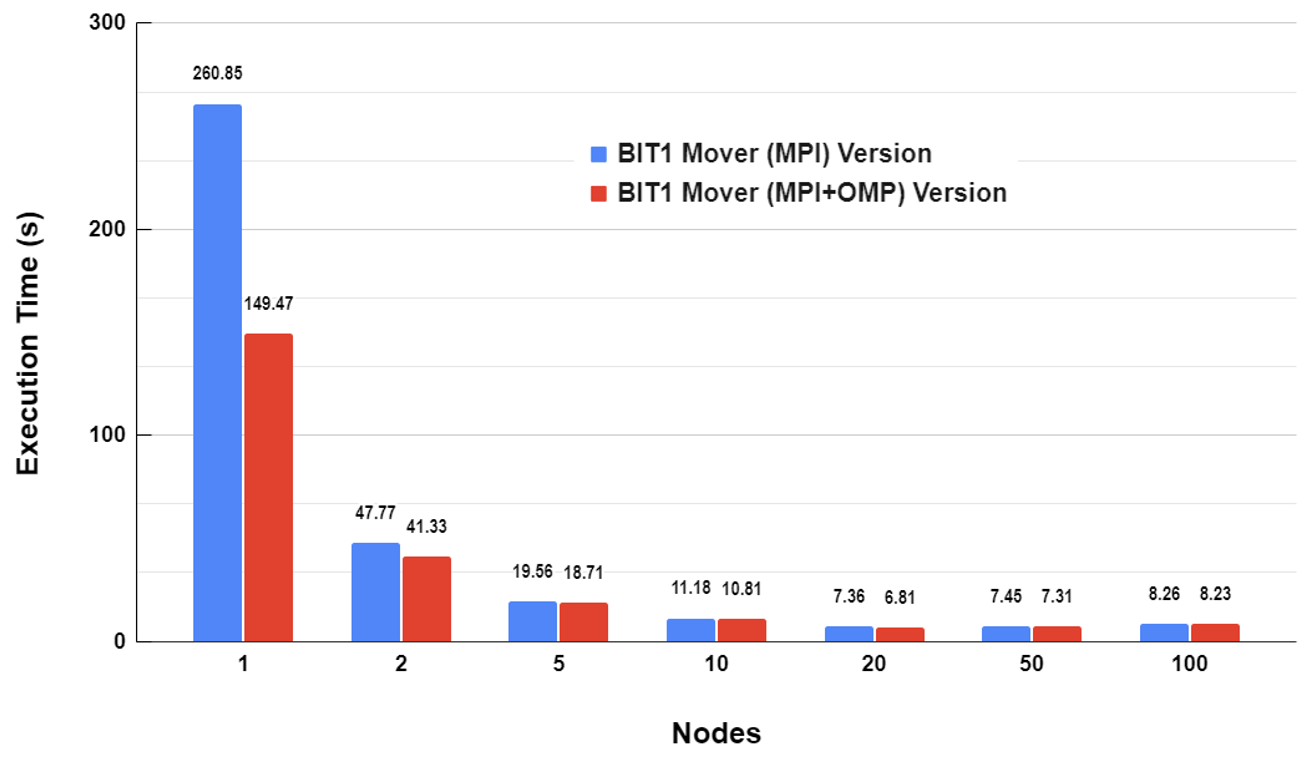}
        \caption{BIT1 optimized mover function - strong scaling up to 100 Nodes (12,800 MPI ranks) on \emph{Dardel} for 200000 times steps.} \label{BIT_Scaling_Dardel_100_Nodes_Mover}
     \end{center}
\end{figure*}
For the BIT1 mover function and Fig.\ref{BIT_Scaling_Dardel_100_Nodes_Mover}, using the MPI version with 1 node yields a time of 260.85 seconds, while the hybrid MPI+OpenMP version improves this to 149.47 seconds, achieving a 42.7\% speedup. As we scale to 10 nodes, the mover function time decreases to 11.18 seconds for MPI and 10.81 seconds for MPI+OpenMP, resulting in a 3.3\% improvement. However, at higher node counts (50 and 100 nodes), the performance remains relatively stable, with times of 7.45 seconds and 8.26 seconds for MPI, and 7.31 seconds and 8.23 seconds for MPI+OpenMP, respectively.

\begin{figure*}[h!]
    \begin{center}
        \includegraphics[width=0.9\textwidth]{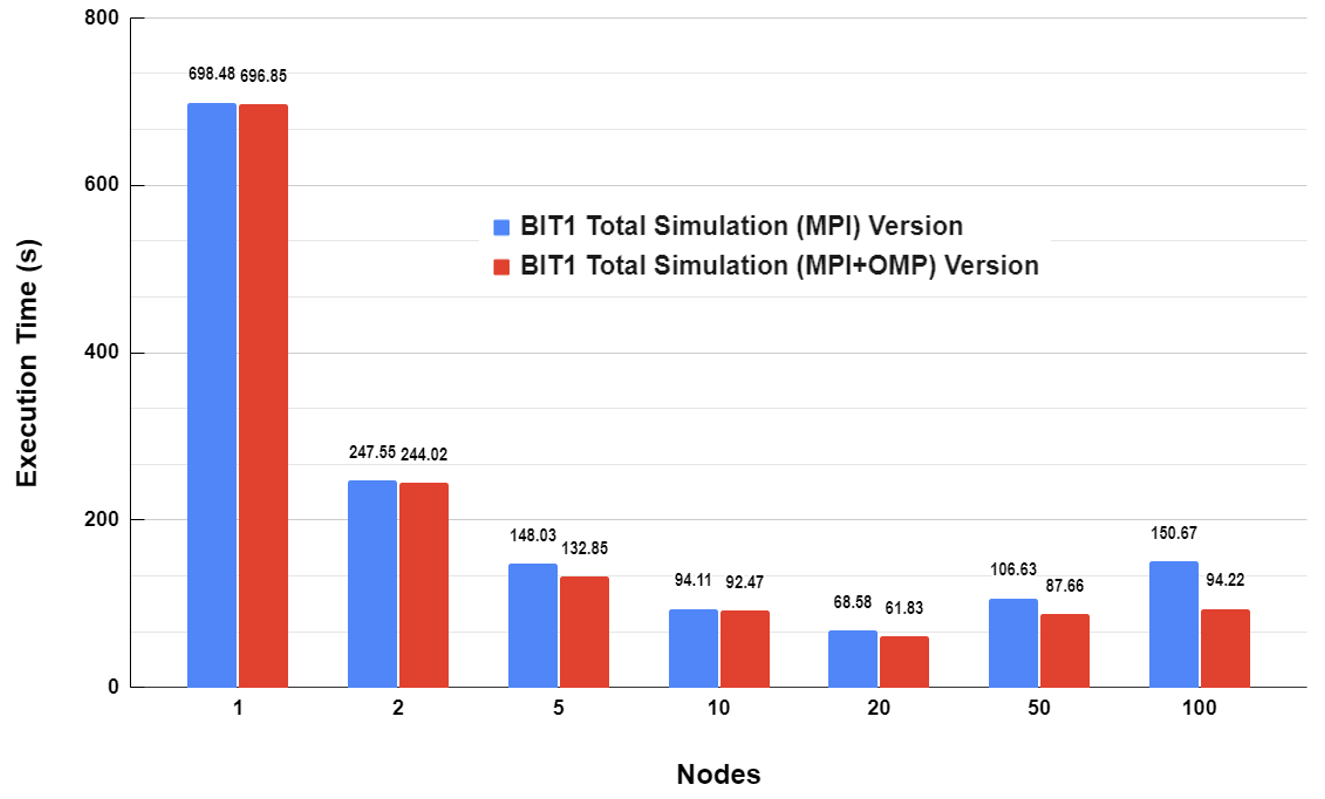}
        \caption{BIT1 Total Simulation - strong scaling up to 100 Nodes (12,800 MPI ranks) on \emph{Dardel} for 200000 times steps.} \label{BIT_Scaling_Dardel_100_Nodes_Total_Simulation}
     \end{center}
\end{figure*}

Similarly, for the BIT1 total simulation and Fig.~\ref{BIT_Scaling_Dardel_100_Nodes_Total_Simulation}, with 1 node, the MPI version completes in 698.48 seconds, while the hybrid MPI+OpenMP version takes 696.85 seconds, resulting in only a negligible improvement. At 2 nodes, the MPI time of 247.55 seconds reduces to 244.02 seconds with MPI+OpenMP, yielding a 1.0\% speedup. The most significant improvements occur at 5 nodes, where the total simulation time drops from 148.03 seconds to 132.85 seconds, achieving a 10.2\% speedup. At 20 nodes, the MPI version takes 68.58 seconds, while the hybrid version further reduces this to 61.83 seconds, resulting in an 11.9\% speedup. However, at 50 nodes, the total simulation time increases for MPI to 106.63 seconds and decreases to 87.66 seconds for MPI+OpenMP, demonstrating a 17.8\% improvement. Finally, at 100 nodes, the total simulation times are 150.67 seconds for MPI and 94.22 seconds for MPI+OpenMP, resulting in an 37.5\% speedup. 

Understanding weak scaling is crucial for evaluating how well a code can handle increasing problem sizes as more computational resources are added. This type of scaling helps assess whether performance improvements are achieved when the problem size per processor remains constant while increasing the number of processors or nodes. 

\begin{figure*}[h!]
    \begin{center}
        \includegraphics[width=0.9\textwidth]{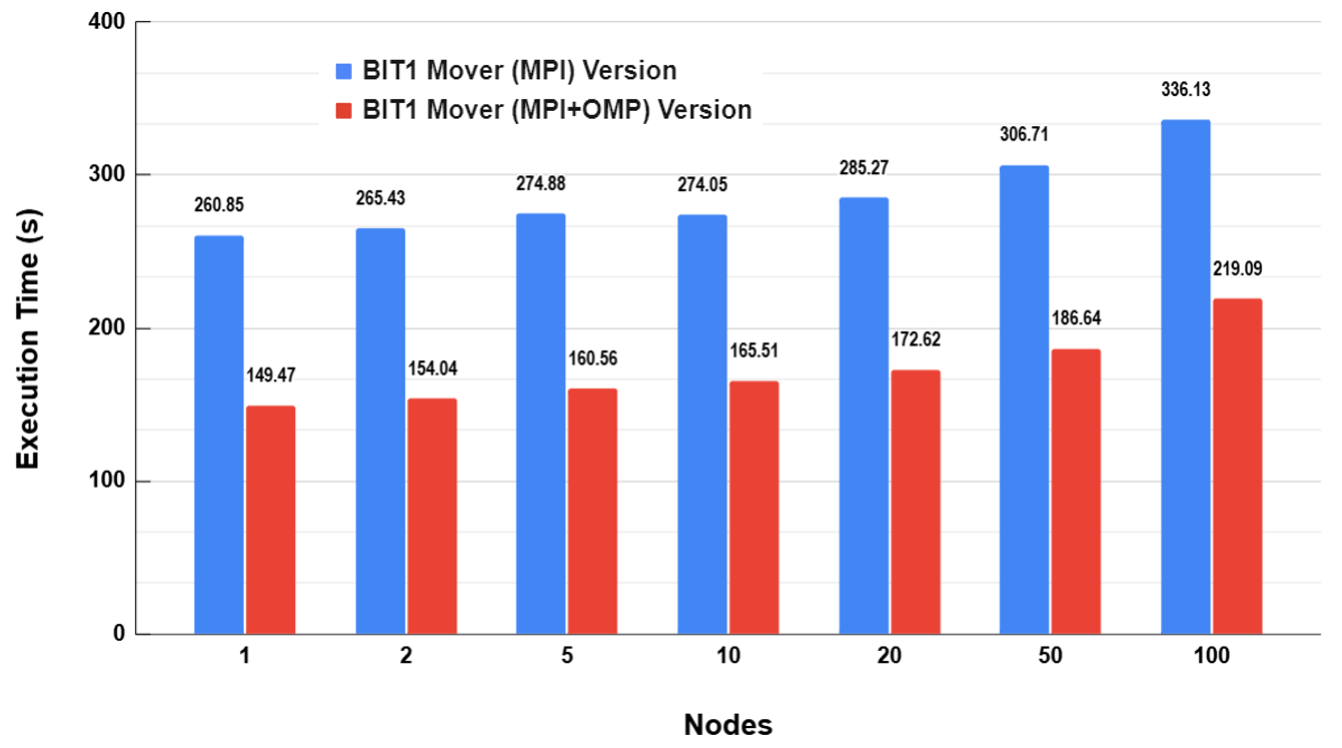}
        \caption{BIT1 optimized mover function - weak scaling up to 100 Nodes (12,800 MPI ranks) on \emph{Dardel} for 200000 times steps.} \label{BIT_Weak_Scaling_Dardel_100_Nodes_Mover}
     \end{center}
\end{figure*}

The weak scaling results for the optimized mover function and the total simulation on \emph{Dardel} provide valuable insights into the performance of the MPI-only and MPI+OMP implementations. For the mover function and Fig.~\ref{BIT_Weak_Scaling_Dardel_100_Nodes_Mover}, the MPI+OMP implementation consistently outperforms the MPI-only version as the number of nodes increases. At 100 nodes, the MPI+OMP version achieves a significant improvement, with a runtime of 219.09 seconds compared to the MPI version’s 336.13 seconds, demonstrating a 34.91\% performance gain. The MPI-only version shows a slight increase in runtime starting from 2 nodes, reaching a peak at 100 nodes. In contrast, the MPI+OMP version exhibits more consistent performance with steady improvements as the number of nodes increases, suggesting better weak scaling for the MPI+OMP parallelization, especially for the mover function. 

\begin{figure*}[h!]
    \begin{center}
        \includegraphics[width=0.8\textwidth]{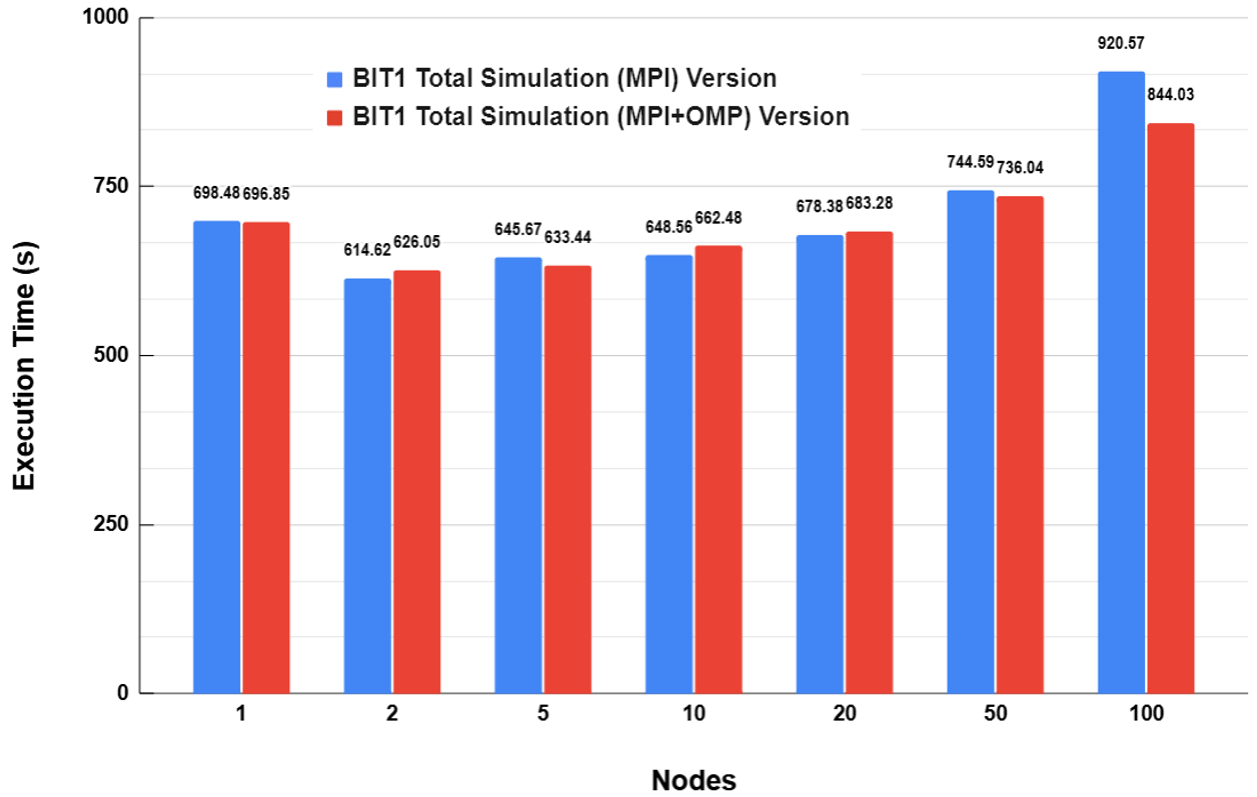}
        \caption{BIT1 Total Simulation - weak scaling up to 100 Nodes (12,800 MPI ranks) on \emph{Dardel} for 200000 times steps.} \label{BIT_Weak_Scaling_Dardel_100_Nodes_Total_Simulation}
     \end{center}
\end{figure*}

For the total simulation and Fig.~\ref{BIT_Weak_Scaling_Dardel_100_Nodes_Total_Simulation}, however, the results show more mixed behavior. While the MPI+OMP version offers stable scalability, it does not always outperform the MPI-only version. At 100 nodes (12,800 MPI ranks), the MPI+OMP version runtime is 844.03 seconds, while the MPI version reaches 920.57 seconds. Despite this, the MPI+OMP version demonstrates more consistent performance improvements at smaller node counts. For instance, at 50 nodes, the MPI version shows a larger increase in runtime (744.59 seconds) compared to the MPI+OMP version (736.04 seconds). 

While the MPI+OMP implementation provides better overall scalability, particularly for the mover function at larger node counts, there is still room for improvement in the full code's performance, especially as the problem size and resource utilization grow. This analysis highlights the benefits of hybrid parallelization in improving both the mover function and total simulation performance on \emph{Dardel}, with further optimizations needed to maximize its potential.

\subsection{Porting BIT1 with 2 GPUs}
The challenge of enhancing the performance of the particle mover function in BIT1 by tapping into the computational power of GPUs has been systematically deliberated, revealing valuable insights. In addressing this challenge, our focus shifts to improving the particle mover function's \texttt{execution time} by offloading it to the GPU using OpenACC and OpenMP. In doing this, we are one step closer for BIT1 being ready for Exascale platforms. To achieve this target, two main strategies provided by OpenACC and OpenMP were investigated: the explicit approach, where data regions for GPU offloading are clearly defined using directives, and the unified memory method, which simplifies memory management by sharing a common space between the CPU and GPU.

\begin{figure*}[h!]
    \begin{center}
        \includegraphics[width=\textwidth]{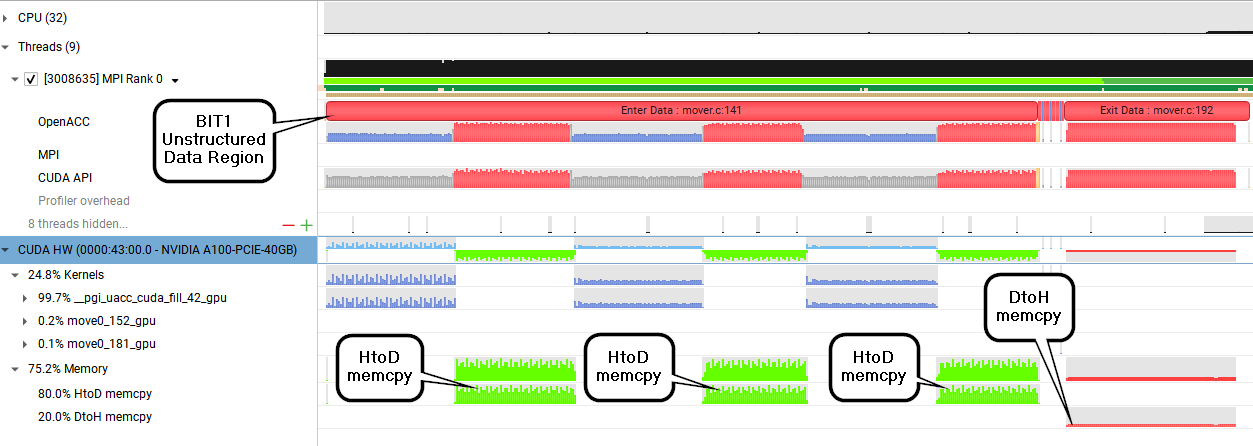}
        \caption{NVIDIA Nsight Systems OpenACC Explicit View - GPU porting of BIT1 mover function on \emph{NJ} with 1 time step.} \label{BIT1_Explicit}
     \end{center}
\end{figure*}

\noindent \textbf{BIT1 OpenACC GPU Explicit.}  Initial work began by using OpenACC’s explicit approach on the GPUs on \emph{NJ} for up to 10 time steps for better visualization of profiling results, particularly focusing on the particle mover function. The profiling results, using NVIDIA Nsight Systems, reveal crucial insights. The CUDA kernel statistics showed that the primary kernel responsible for the particle mover function consumes 99.7\% of the GPU execution time, emphasizing its significance in the overall computation. In Fig.~\ref{BIT1_Explicit}, the memory operation statistics shed light on the substantial time spent on \verb|memcpy| operations. Specifically, 80\% of the GPU time is allocated to copying data from host to device, emphasizing the importance of efficient data transfer strategies.

These findings emphasize the importance of optimizing data movement and kernel execution in the particle mover function. Strategies such as overlapping computation and communication, along with exploring ways to minimize data transfer size, can be instrumental. Additionally, considering the high memory bandwidth of the NVIDIA A100 GPUs on \emph{NJ}, enhancing the efficiency of memory operations becomes pivotal for achieving optimal performance.

\noindent \textbf{BIT1 OpenACC GPU Unified Memory.} Next, the particle mover function has been initially investigated on \emph{NJ} using OpenACC Unified Memory. The profiling results revealed critical insights into the dynamics of memory management and kernel execution. 

The CUDA kernel statistics showed the primary mover kernel $move0_{152\_gpu}$ dominated GPU execution time, accounting for 66.7\% with a total execution time of 1.623457 seconds. Similarly the second kernel $move0_{181\_gpu}$ contributed 33.3\% with an average execution time of 0.8106292 seconds, highlighting the significance of these kernels in the overall computation. 

\begin{figure*}[h!]
    \begin{center}
        \includegraphics[width=\textwidth]{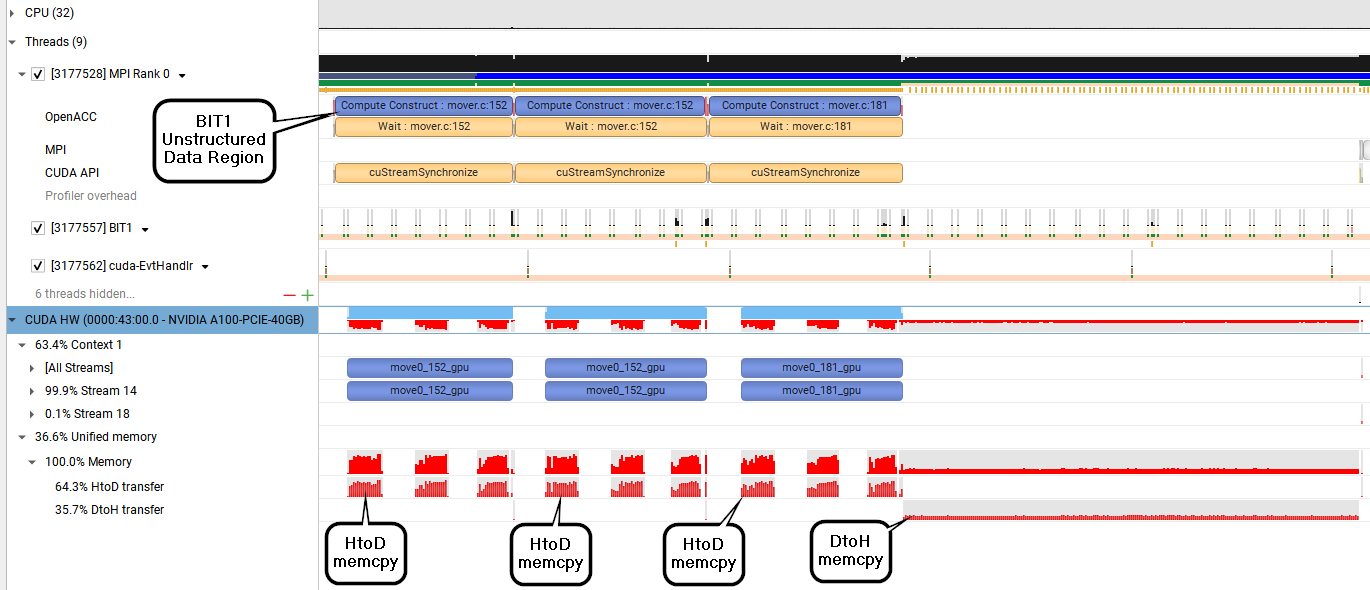}
        \caption{NVIDIA Nsight Systems OpenACC Unified Memory View - GPU porting of BIT1 mover function on \emph{NJ} with 1 time step.} \label{BIT1_Unified_Memory}
     \end{center}
\end{figure*}

For hybrid BIT1 data movement and the implementation of a unified memory strategy, the runtime system manages the seamless transfer of data between the host and the device. One key advantage, aside from programmer convenience, is the opportunity for the runtime to automatically detect instances of overlapping computation and communication. Fig.~\ref{BIT1_Unified_Memory} displays NVIDIA Nsight Systems' view of such overlapping, revealing that the unified memory version exhibited faster overall runtime than explicit copies. This indicates automatic overlapping of computation and communication. However, we observe that BIT1 OpenACC Unified Memory performance is still hindered by substantial data movement between the host and device.

\begin{figure*}[h!]
    \begin{center}
        \includegraphics[width=0.8\textwidth]{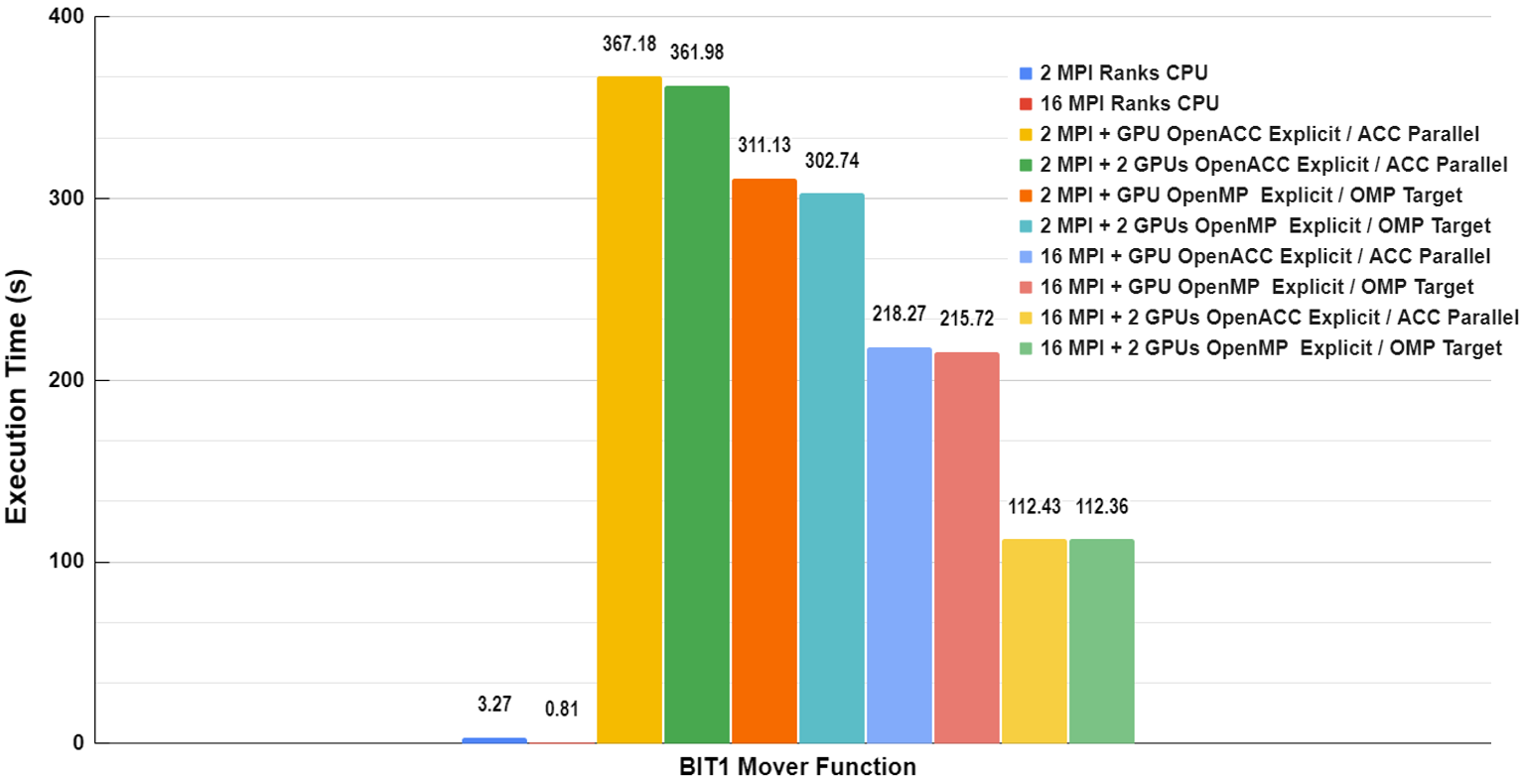}
        \caption{Optimized mover function execution time(s) on \emph{NJ} for 100 time steps using OpenMP and OpenACC Explicit on NVIDIA GPUs.} \label{BIT_GPU_Explicit_Mover_Execution_Time}
     \end{center}
\end{figure*}
\begin{figure*}[h!]
    \begin{center}
        \includegraphics[width=0.8\textwidth]{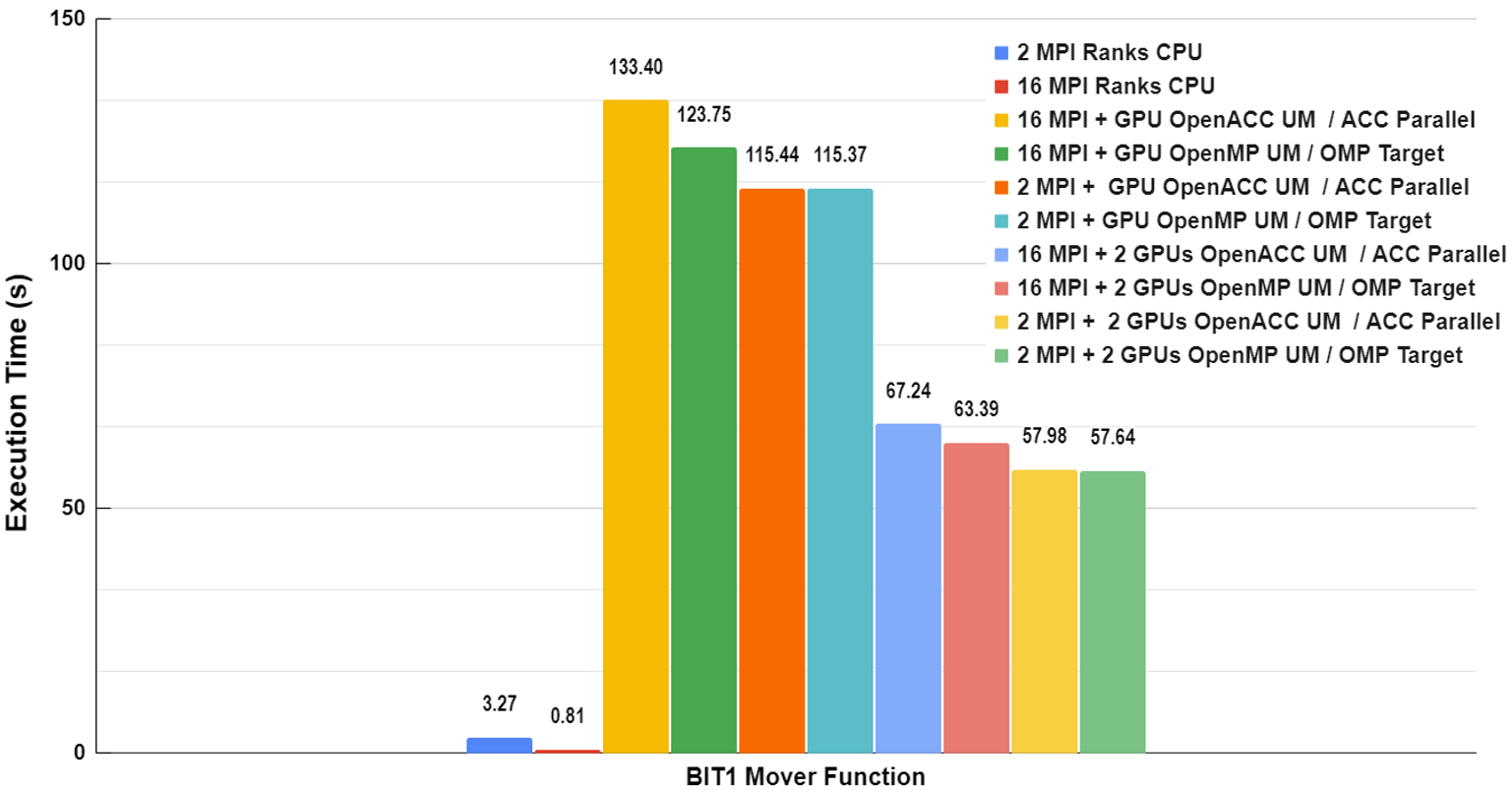}
        \caption{Optimized mover function execution time(s) on \emph{NJ} for 100 time steps using OpenMP and OpenACC Unified Memory on NVIDIA GPUs.} \label{BIT_GPU_Unified_Memory_Mover_Execution_Time}
     \end{center}
\end{figure*}

\noindent \textbf{BIT1 OpenMP GPU Offloading.} Due to the observed performance gain with hybrid BIT1 on multicore CPU, an investigation was conducted using the OpenMP target construct for further BIT1 GPU offloading. Fig.~\ref{BIT_GPU_Explicit_Mover_Execution_Time} and~\ref{BIT_GPU_Unified_Memory_Mover_Execution_Time} provides performance evaluation insights into GPU utilization compared to the CPU-only baseline with 2 MPI and 16 MPI Ranks. Implementations using OpenMP or OpenACC on GPUs exhibit increased execution times, indicating that parallelization strategies may introduce overhead, potentially outweighing the benefits of parallel processing. Among BIT1 GPU implementations, OpenMP Target with 2 GPUs stands out for delivering a substantial reduction in execution time, demonstrating the performance improvement through concurrent GPU utilization, especially when MPI ranks are assigned to dedicated GPUs.

\subsection{Scaling BIT1 with 64 GPUs}
Continuing our initial efforts in porting BIT1 with 2 GPUs, we study the scalability of enhancing the particle mover function with up to 64 GPUs. This helps us understand key aspects of GPU utilization, data movement, and bottlenecks that could impact performance at large scales.

\begin{figure*}[h!]
    \begin{center}
        \includegraphics[width=\textwidth]{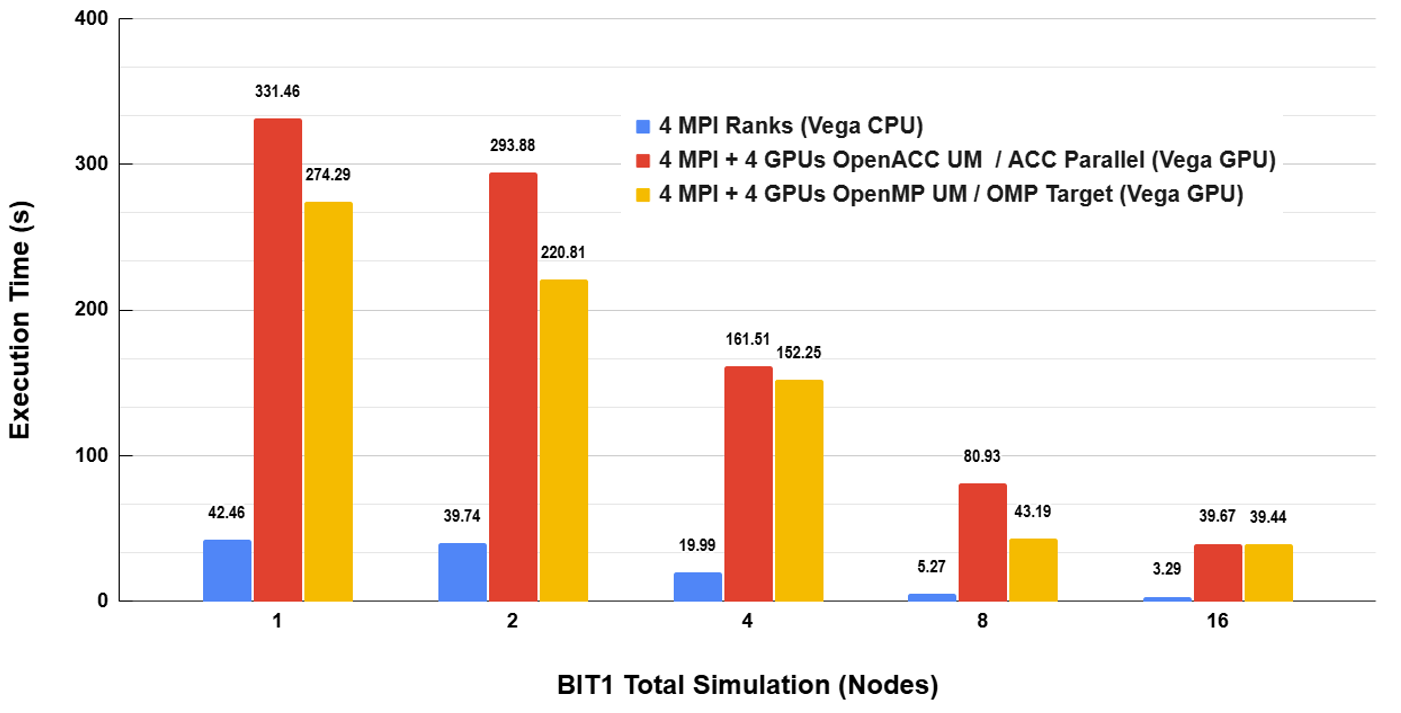}
        \caption{BIT1 Total Simulation (strong scaling) execution time(s) on \emph{Vega} for 200 time steps up to 16 Nodes (64 MPI Ranks and 64 GPUs) using OpenMP and OpenACC Unified Memory on NVIDIA GPUs.} \label{BIT_GPU_Unified_Memory_Total_Execution_Time_Vega}
     \end{center}
\end{figure*}

\begin{figure*}[h!]
    \begin{center}
        \includegraphics[width=\textwidth]{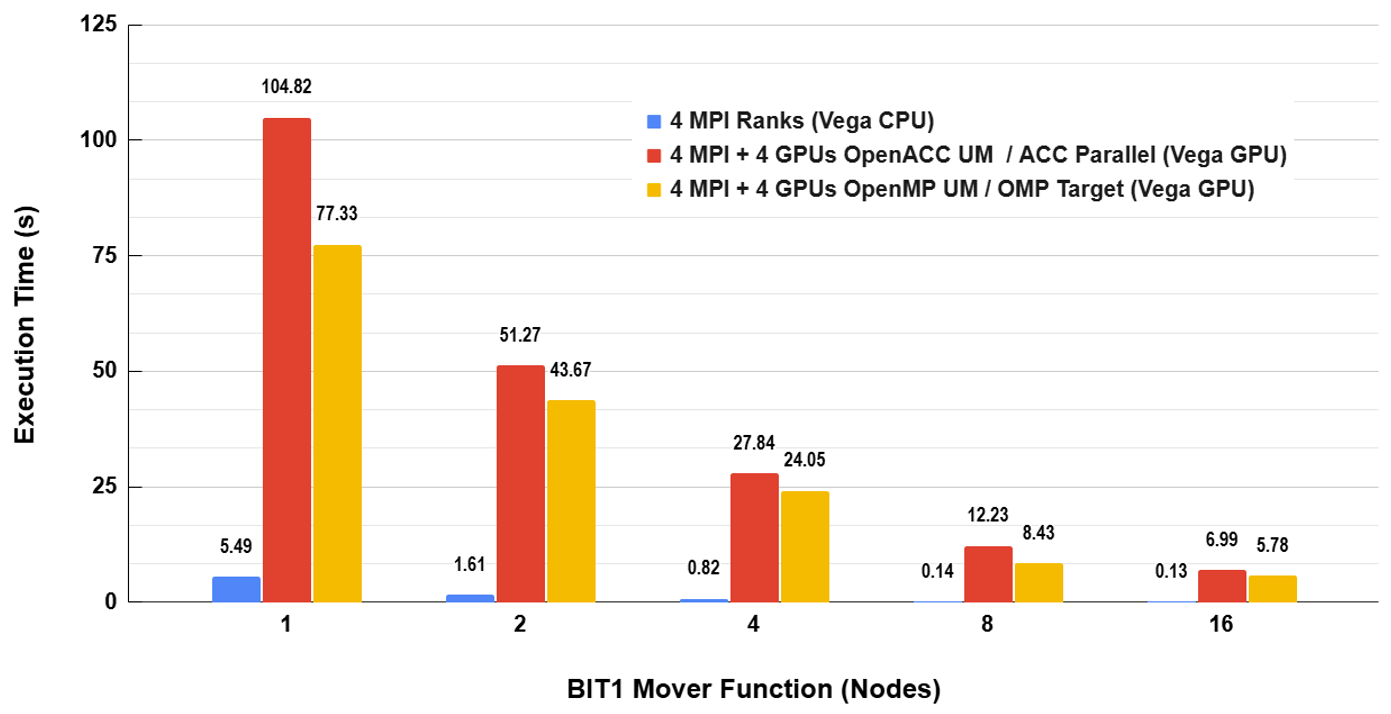}
        \caption{BIT1 Optimized mover function (strong scaling) execution time(s) on \emph{Vega} for 200 time steps up to 16 Nodes (64 MPI Ranks and 64 GPUs) using OpenMP and OpenACC Unified Memory on NVIDIA GPUs.} \label{BIT_GPU_Unified_Memory_Mover_Execution_Time_Vega}
     \end{center}
\end{figure*}

\noindent \textbf{Scaling BIT1 w/ 64 GPUs on Vega GPU.} We begin by running BIT1 on Vega, a petascale EuroHPC supercomputer, for up to 16 nodes and 200 timesteps, where each MPI rank is assigned to dedicated GPUs. The performance of BIT1's total simulation and mover function reveals an improvement through GPU acceleration, although the CPU version remains the fastest overall.

For the BIT1 total simulation and in Fig.~\ref{BIT_GPU_Unified_Memory_Total_Execution_Time_Vega}, the slowest GPU version (4 MPI + 4 GPUs OpenACC UM / ACC Parallel) took 331.46 seconds on 1 node, while the fastest GPU version (4 MPI + 4 GPUs OpenMP UM / OMP Target) reduced this to 274.29 seconds, a 17.26\% improvement. On 4 nodes, the fastest BIT1 GPU configuration achieved a 5.73\% speedup (from 161.51 to 152.25 seconds), and at 16 nodes, the speedup was minimal (from 39.67 to 39.44 seconds). Despite these gains, the CPU version remained faster across all nodes, taking only 42.46 seconds on 1 node and 3.29 seconds on 16 nodes, consistently outperforming even the fastest GPU versions.

For the mover function and in Fig.~\ref{BIT_GPU_Unified_Memory_Mover_Execution_Time_Vega}, the BIT1 4 MPI + 4 GPUs OpenMP UM / OMP Target configuration also outpaced the standard OpenACC version, reducing execution time by 26.15\% on 1 node (from 104.82 to 77.33 seconds). On 4 nodes, it showed a 15.65\% speedup (from 27.84 to 24.05 seconds), and a 20.47\% improvement at 16 nodes (from 6.99 to 5.78 seconds). However, the CPU version was still faster, completing the mover in just 0.13 seconds at 16 nodes compared to 5.78 seconds for the fastest GPU configuration. 

\subsection{Accelerating BIT1 Asynchronously with 128 GPUs}
Extending from our initial efforts in porting BIT1 with 64 GPUs, we now enhance the particle mover function by implementing the first asynchronous multi-GPU version with up to 128 GPUs. This advancement leverages OpenMP Target Tasks with the "nowait" and "depend" clauses and OpenACC parallel regions using the "async(n)" clause to optimize \texttt{execution time} through concurrent executions. Additionally, we will focus our efforts on using Unified Memory (UM), which simplifies memory management by sharing a common space between the CPU and GPU, as it has demonstrated faster execution compared to the explicit method. This asynchronous approach improves load balancing and resource utilization, positioning BIT1 to exploit Exascale platforms effectively. 

\begin{figure*}[h!]
    \begin{center}
        \includegraphics[width=\textwidth]{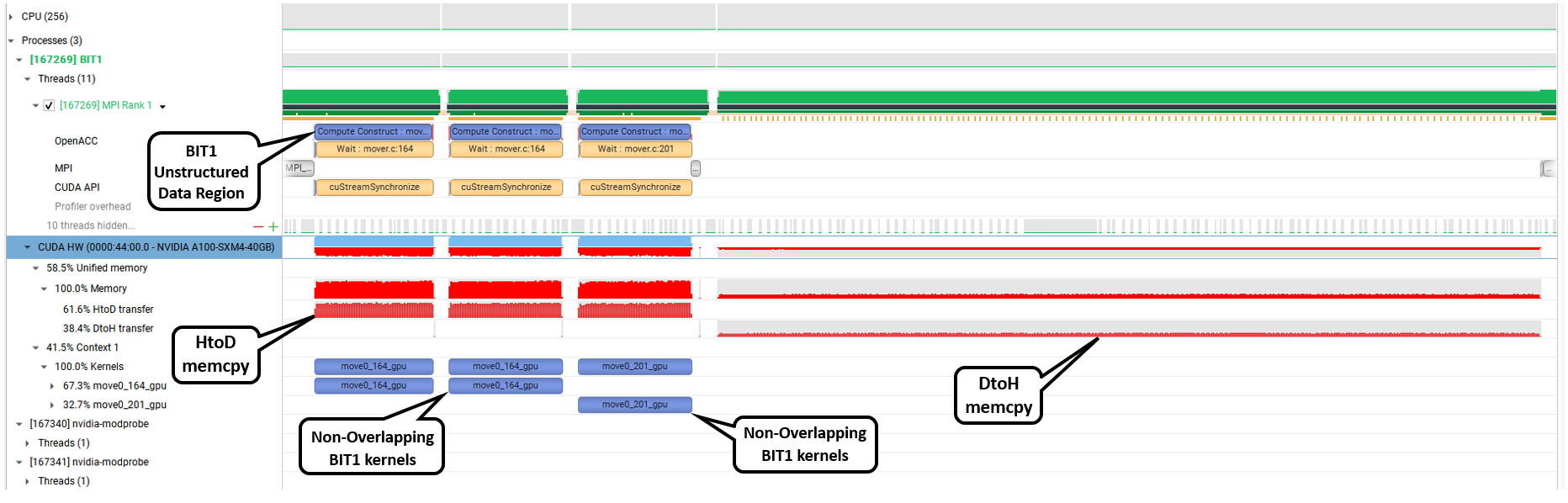}
        \caption{NVIDIA Nsight Systems OpenACC UM / ACC Parallel View - GPU porting of BIT1 mover function on \emph{Vega} with 1 time step.} 
        \label{BIT1_Profiling_Vega} 
     \end{center}
\end{figure*}

\begin{figure*}[h!]
    \begin{center}
        \includegraphics[width=\textwidth]{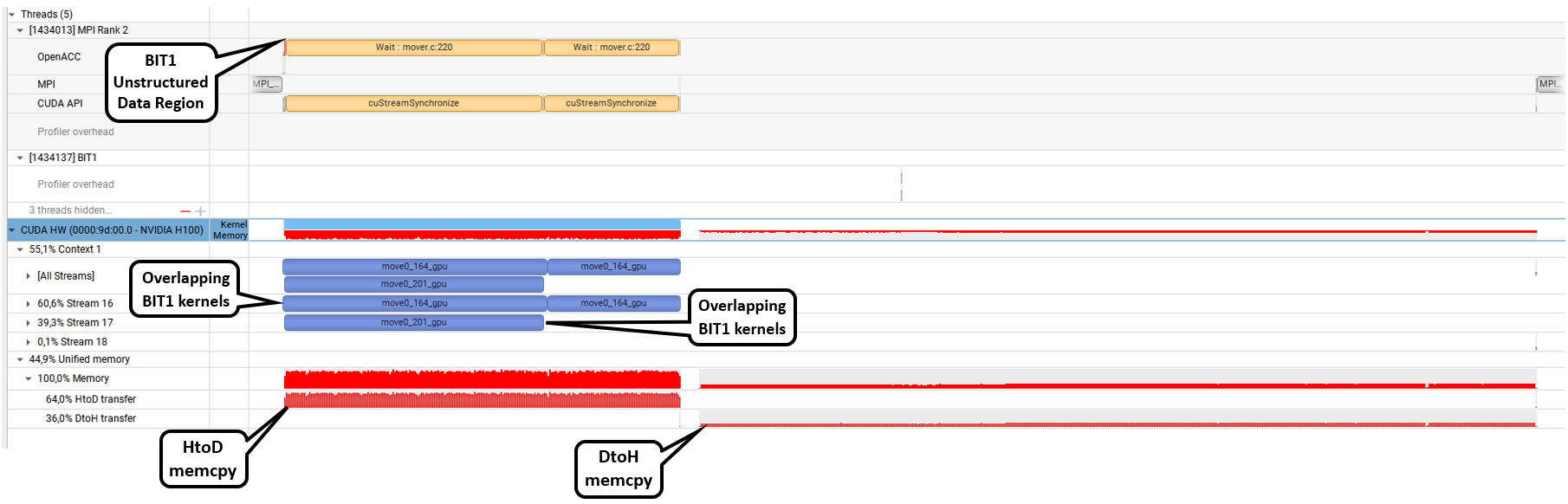}
        \caption{NVIDIA Nsight Systems OpenACC UM / ACC Parallel Async(n) View - GPU porting of BIT1 mover function on \emph{MN5} with 1 time step.} \label{BIT1_Async_Profiling_MN5}
     \end{center}
\end{figure*}

\noindent \textbf{Accelerating BIT1 Asynchronously on MN5 ACC}
We critically analyze and discuss the performance of the accelerated BIT1 code on a pre-exascale EuroHPC supercomputer, \emph{MN5 H100} NVIDIA ACC partition, scaling up to 32 nodes with 200 timesteps. This serves as an initial step in preparing for the upcoming EuroHPC exascale supercomputer. As in previous tests, each configuration utilized 4 MPI ranks, with each GPU version having a dedicated GPU assigned to each MPI rank. This avoids any competition for GPU resources.

Using NVIDIA Nsight Systems, we reveal crucial insights into the porting efforts and acceleration of the particle mover function on GPUs. Previously, on \emph{NJ}, as shown in Fig.~\ref{BIT1_Unified_Memory}, and now on \emph{Vega} in Fig. ~\ref{BIT1_Profiling_Vega}, the primary kernel consumed around 99.7\% of the GPU execution time for both \emph{NJ} and \emph{Vega} without overlapping computation and communication. Now, \emph{MN5} displays overlapping kernels, with data transfers and computations managed together for accurate results without blocking the CPU, thereby improving execution time in the mover function. This can be seen in Fig.~\ref{BIT1_Async_Profiling_MN5}, where the primary kernel consumption has been reduced to 58.6\% due to overlapping computation and communication on \emph{MN5}.

\begin{figure*}[h!]
    \begin{center}
        \includegraphics[width=\textwidth]{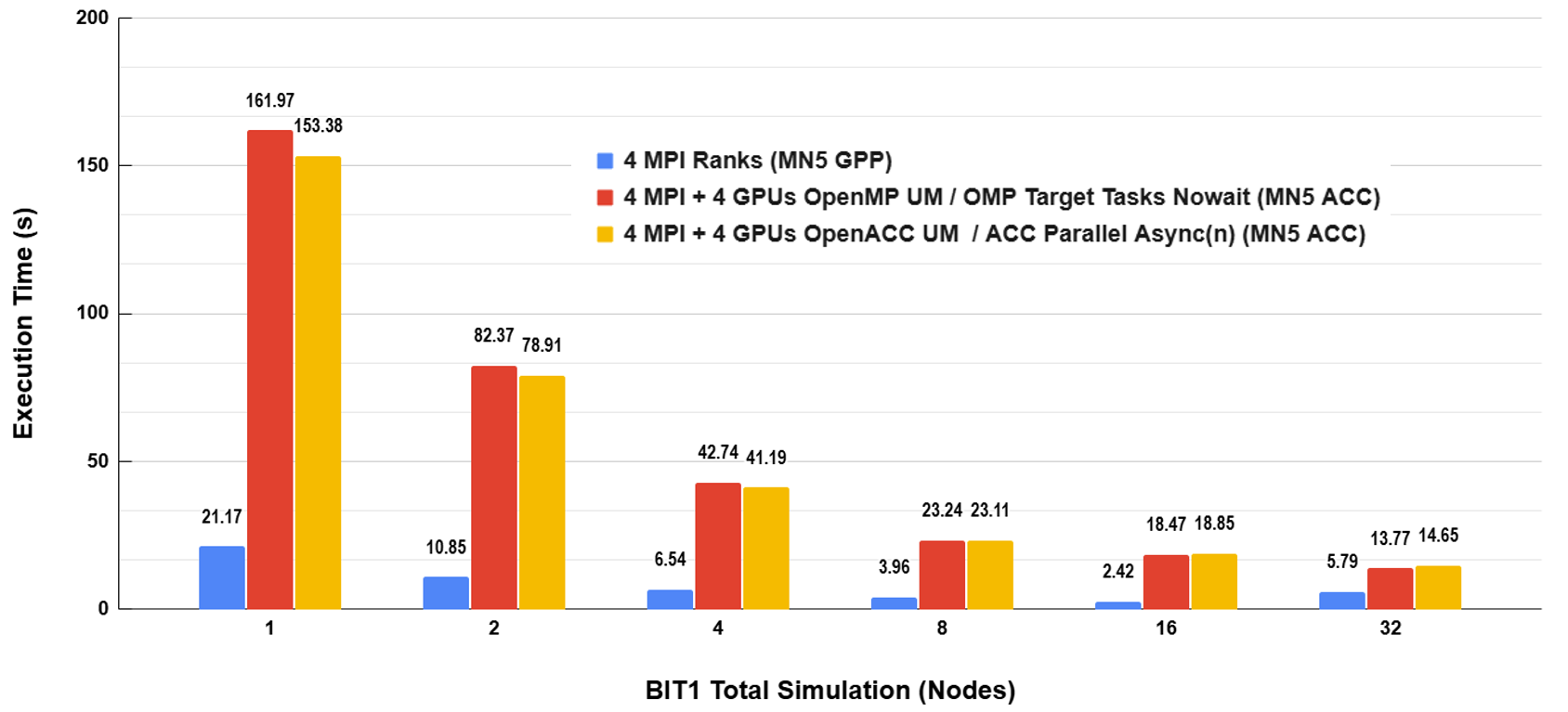}
        \caption{BIT1 Total Simulation (strong scaling) execution time(s) for 200 time steps up to 32 Nodes (128 MPI Ranks and 128 GPUs) using OpenMP and OpenACC Unified Memory on \emph{MN5 H100} NVIDIA GPUs.} \label{BIT_GPU_Unified_Memory_Total_Simulation_Execution_Time_MN5}
     \end{center}
\end{figure*} 

\begin{figure*}[h!]
    \begin{center}
        \includegraphics[width=\textwidth]{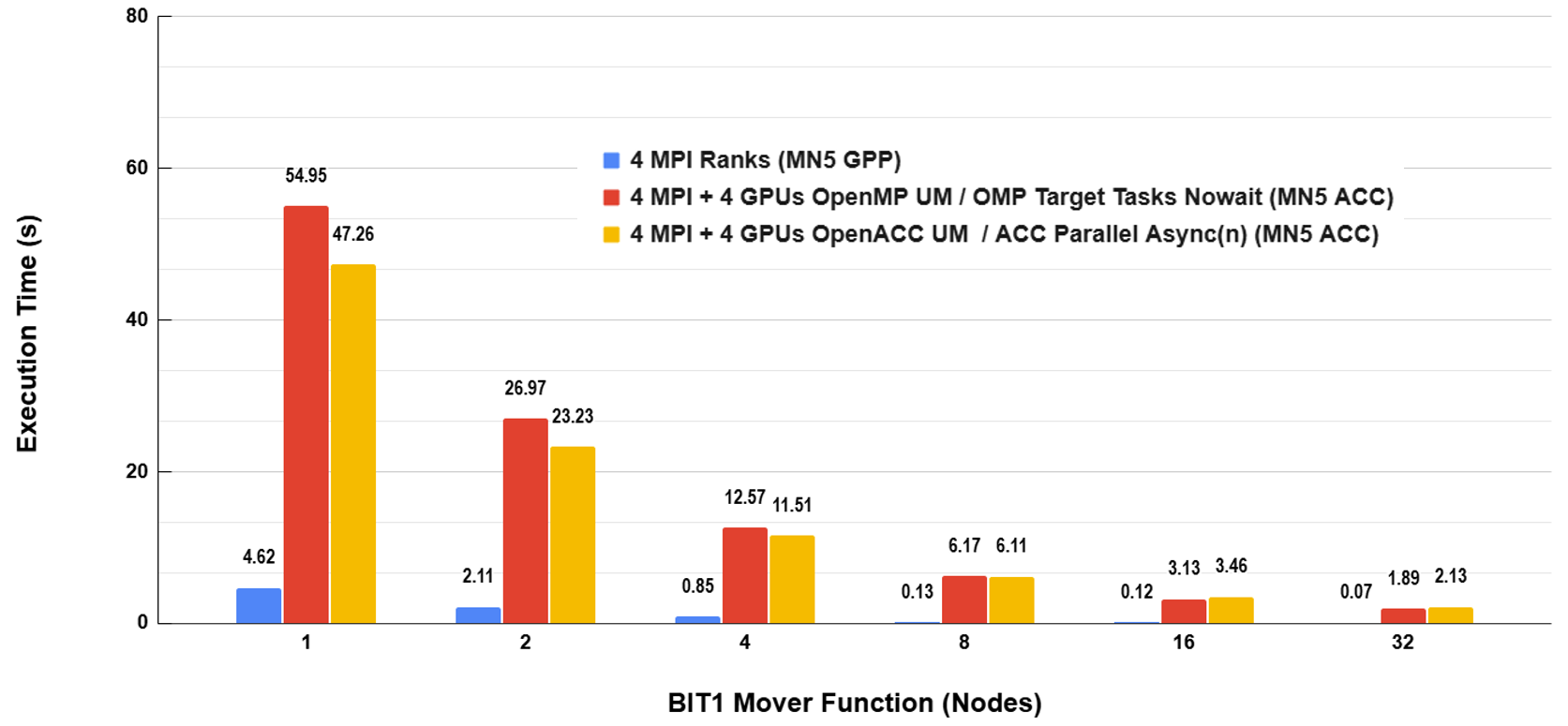}
        \caption{BIT1 Optimized mover function (strong scaling)  execution time(s) for 200 time steps up to 32 Nodes (128 MPI Ranks and 128 GPUs) using OpenMP and OpenACC Unified Memory on \emph{MN5 H100} NVIDIA GPUs.} \label{BIT_GPU_Unified_Memory_Mover_Execution_Time_MN5}
     \end{center}
\end{figure*} 

For the total simulation and as shown in Fig.~\ref{BIT_GPU_Unified_Memory_Total_Simulation_Execution_Time_MN5}, the slowest GPU version (OpenMP Nowait) took 161.97 seconds on 1 node, while the fastest GPU version (OpenACC Async(n)) reduced this to 153.38 seconds, achieving a 5.31\% improvement. At 4 nodes, the OpenACC Async(n) version showed a 3.63\% speedup over the OpenMP Nowait version (from 42.74 to 41.19 seconds). Between 8 and 16 nodes, OpenACC Async(n) version demonstrated a slight advantage, with a 0.56\% improvement at 8 nodes and a marginal 2.05\% slowdown at 16 nodes. At 32 nodes, the OpenACC Async(n) version showed an execution time of 14.65 seconds, 6.39\% slower than the OpenMP Nowait version, indicating possible saturation effects and reduced scalability. Despite these gains, the CPU version consistently outperformed the GPU configurations, taking only 21.17 seconds on 1 node and 2.42 seconds on 16 nodes, but slowing down to 5.79 seconds at 32 nodes, revealing a similar saturation trend. 

For the mover function, as shown in Fig.~\ref{BIT_GPU_Unified_Memory_Mover_Execution_Time_MN5}, the OpenACC Async(n) version outpaced the OpenMP Nowait version, reducing execution time by 14.01\% on 1 node (from 54.95 to 47.26 seconds). At 4 nodes, the OpenACC Async(n) version was 8.45\% faster. However, at 8 nodes, the OpenACC Async(n) version showed a slight slowdown at 6.11 seconds compared to the OpenMP Nowait version at 6.17 seconds. At 16 nodes, it further slowed down to 3.46 seconds compared to the OpenMP Nowait version at 3.13 seconds, and the trend continued at 32 nodes, with the OpenMP Nowait version performing better at 1.89 seconds. Between 8 and 32 nodes, OpenACC Async(n) initially maintained its advantage but then faced scalability challenges, leading to increased execution times, while OpenMP Nowait improved its performance after 16 nodes due to better communication efficiencies and data movement. Similar to the total simulation, the CPU version outperformed the GPU configurations, executing in only 4.62 seconds on 1 node and 0.07 seconds on 32 nodes, maintaining its dominance due to single-threaded performance despite showing saturation at higher node counts for the total simulation.

\begin{figure*}[h!]
    \begin{center}
        \includegraphics[width=\textwidth]{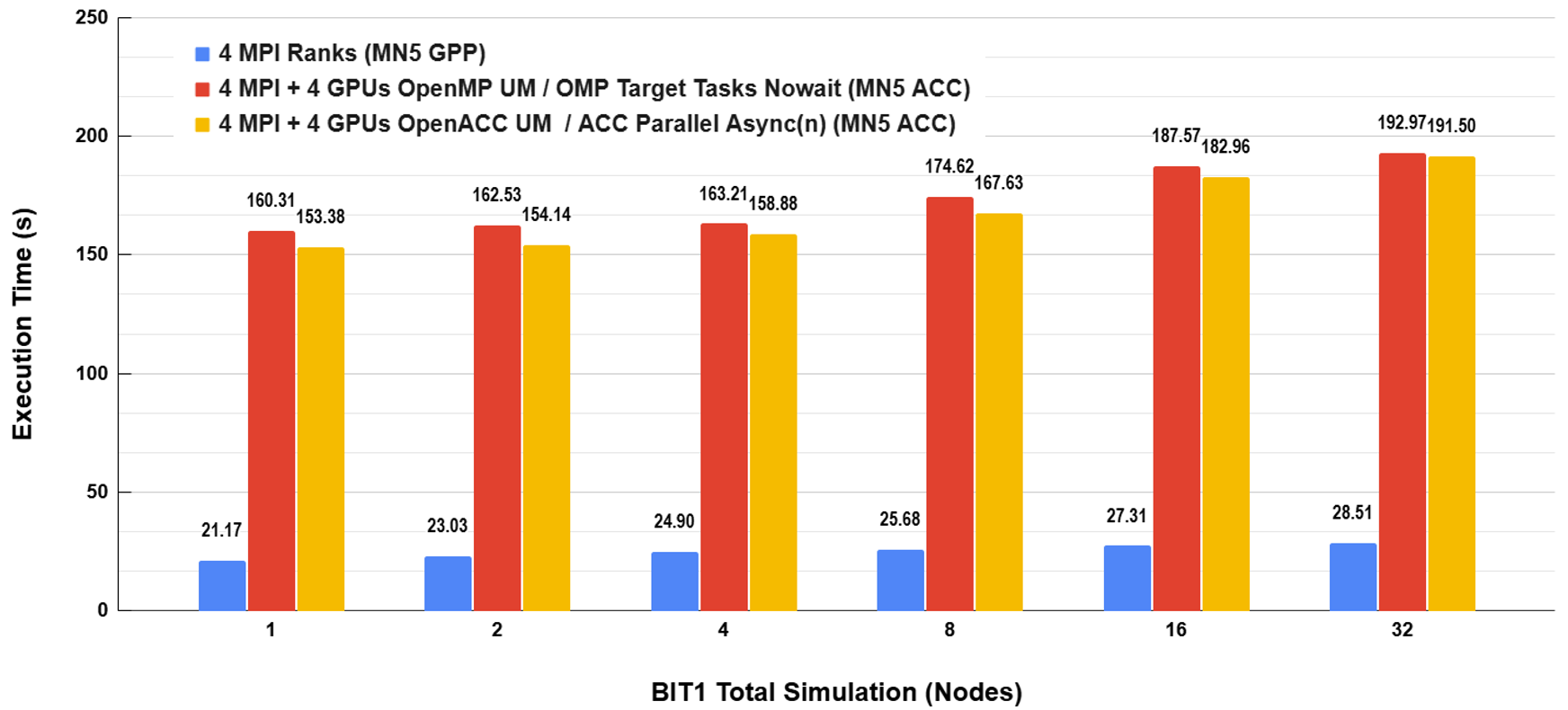}
        \caption{BIT1 Total Simulation (weak scaling) execution time(s) for 200 time steps up to 32 Nodes (128 MPI Ranks and 128 GPUs) using OpenMP and OpenACC Unified Memory on \emph{MN5 H100} NVIDIA GPUs.} 
        \label{BIT_GPU_Unified_Memory_Total_Execution_Time_MN5_Weak}
     \end{center}
\end{figure*}

\begin{figure*}[h!]
    \begin{center}
        \includegraphics[width=\textwidth]{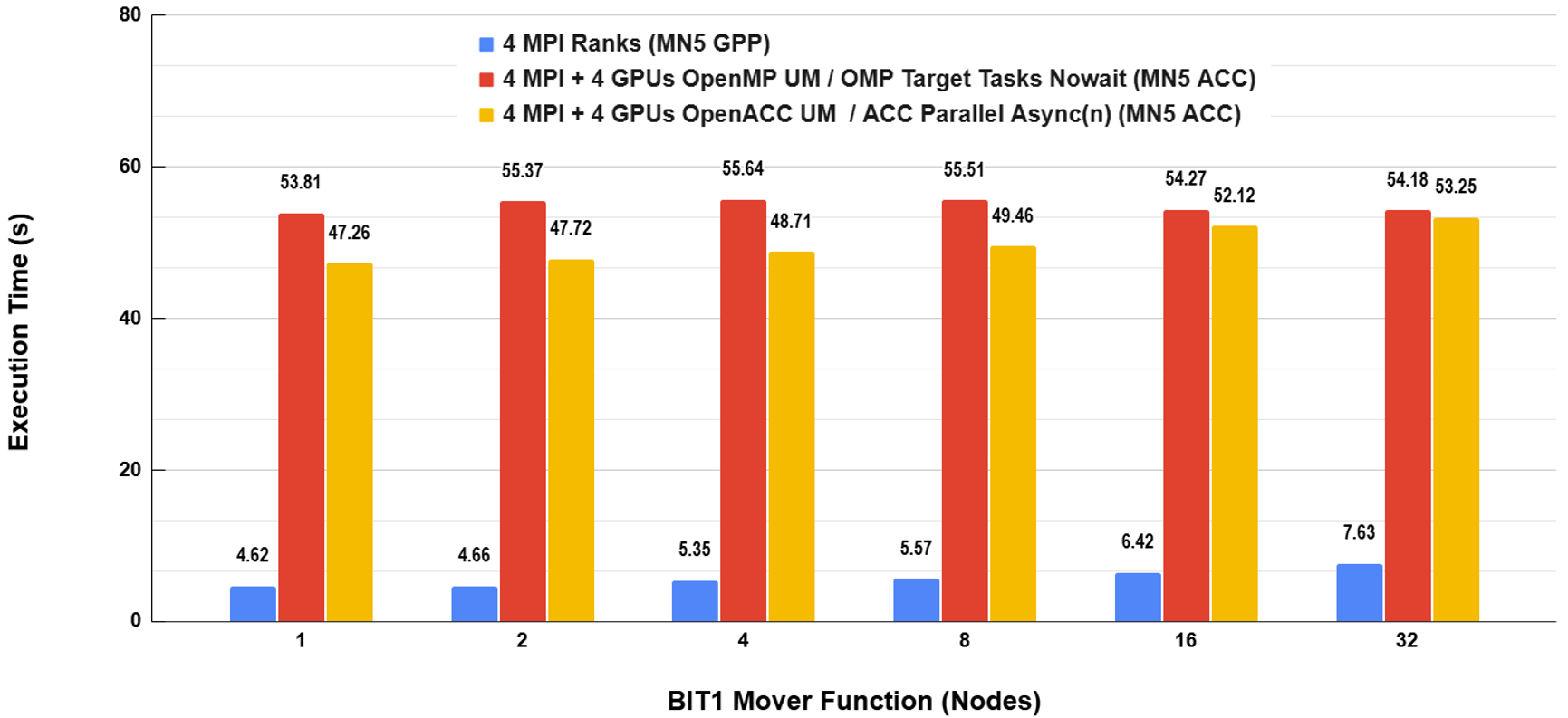}
        \caption{BIT1 Optimized mover function (weak scaling) execution time(s) for 200 time steps up to 32 Nodes (128 MPI Ranks and 128 GPUs) using OpenMP and OpenACC Unified Memory on \emph{MN5 H100} NVIDIA GPUs.} 
        \label{BIT_GPU_Unified_Memory_Mover_Execution_Time_MN5_Weak}
     \end{center}
\end{figure*} 

These results highlight the advantages of GPU acceleration for BIT1 while also indicating that the CPU version remains highly efficient. The observed performance at 8 to 32 nodes suggests diminishing returns from GPU scaling due to communication overhead and data movement bottlenecks, emphasizing the need for further optimization to fully leverage GPU hardware on large-scale systems.

Studying the performance of increasing the problem size and computational resources on GPUs shows interesting results on MN5. For the total simulation and in Fig.~\ref{BIT_GPU_Unified_Memory_Total_Execution_Time_MN5_Weak}, the MN5 GPP (CPU) and ACC (GPU) versions' runtimes increase from 1 node (4 GPUs) to 32 nodes (128 GPUs). These upward trends indicate that none of the versions achieve perfect weak scaling, as the runtimes do not remain constant when scaling both the problem size and computational resources. However, for the mover function offloaded to the GPU and in Fig.~\ref{BIT_GPU_Unified_Memory_Mover_Execution_Time_MN5_Weak}, the performance is more stable. The OpenACC Async(n) version, for example, shows only a slight increase from 47.26 seconds on 1 node to 53.25 seconds on 32 nodes, indicating good weak scaling behavior. Meanwhile, the OpenMP Nowait version demonstrates better weak scaling, with runtimes increasing from 53.81 seconds on 1 node to 55.64 seconds on 4 nodes, followed by a continuous decrease on 8, 16, and 32 nodes (55.51, 54.27, and 54.18 seconds, respectively). Although the OpenMP version's overall runtime is higher compared to OpenACC, its better weak scaling suggests effective load balancing and efficient communication management as the problem size and GPU resources increase despite the slightly higher computational cost.

It's important to note that achieving ideal weak scaling is challenging due to factors such as communication overhead and synchronization costs, which tend to increase with the number of nodes. The relatively stable performance of the GPU versions in the mover function suggests that, with further optimization, offloading computations to GPUs could be promising for large-scale PIC simulations. 

\subsection{Extreme Scaling BIT1 Asynchronously with 400 GPUs}
Investigating further, at extreme scales on \emph{MN5}, the performance analysis of BIT1 highlights both improvements and limitations when scaling up to 400 GPUs. The OpenMP Nowait version demonstrates better overall scaling efficiency compared to OpenACC Async(n), particularly at large GPU counts, as shown in Fig.~\ref{BIT_GPU_Unified_Memory_Total_Simulation_Execution_Time_MN5} and~\ref{BIT_GPU_Unified_Memory_Mover_Execution_Time_MN5}. While both GPU implementations reduce execution time as computational resources increase, OpenMP Nowait consistently outperforms OpenACC at extreme node counts. 

\begin{figure*}[h!]
    \begin{center}
        \includegraphics[width=\textwidth]{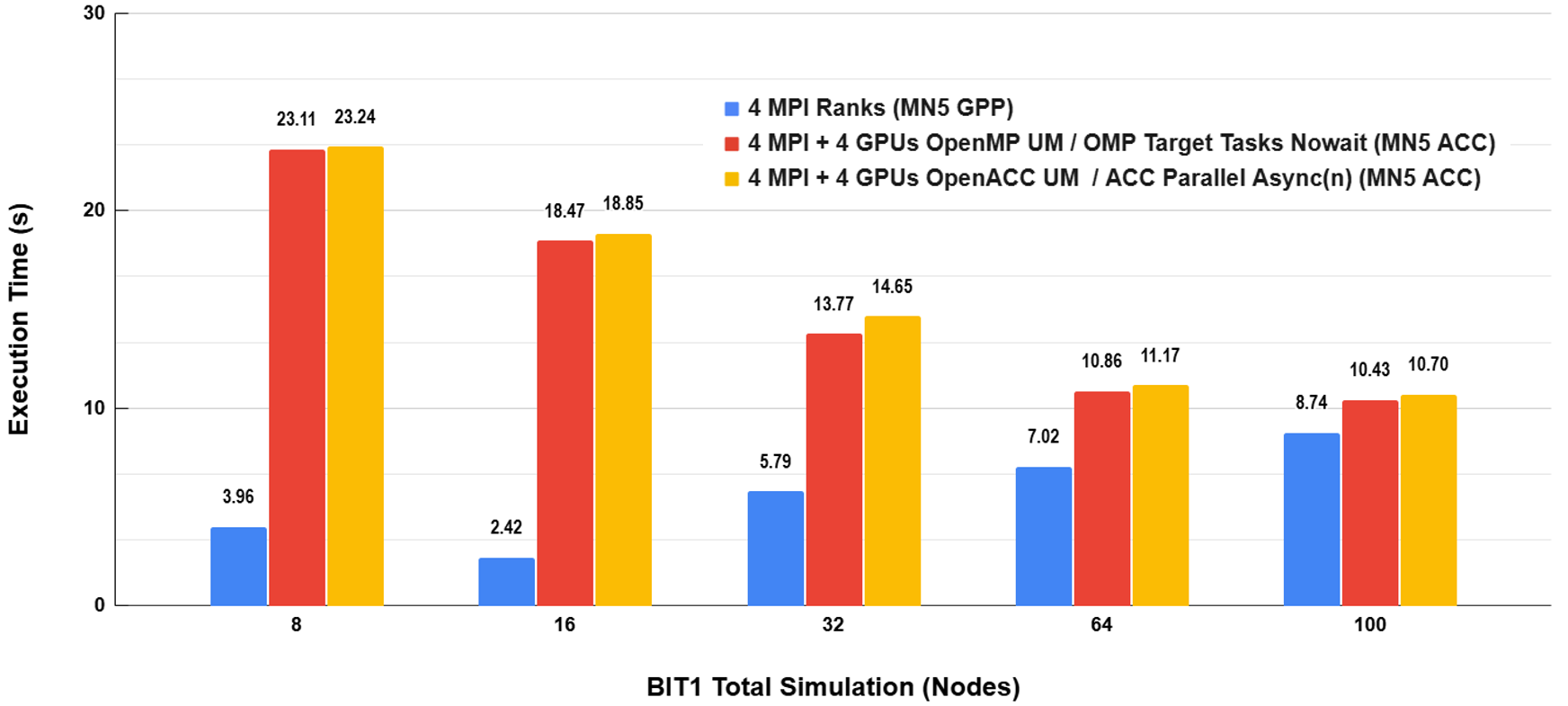}
        \caption{BIT1 Total Simulation (strong scaling) execution time(s) for 200 time steps up to 100 Nodes (400 MPI Ranks and 400 GPUs)  using OpenMP and OpenACC Unified Memory on \emph{MN5 H100} NVIDIA GPUs.} \label{BIT_GPU_Unified_Memory_Total_Simulation_Execution_Time_MN5_Extreme}
     \end{center}
\end{figure*}

\begin{figure*}[h!]
    \begin{center}
        \includegraphics[width=\textwidth]{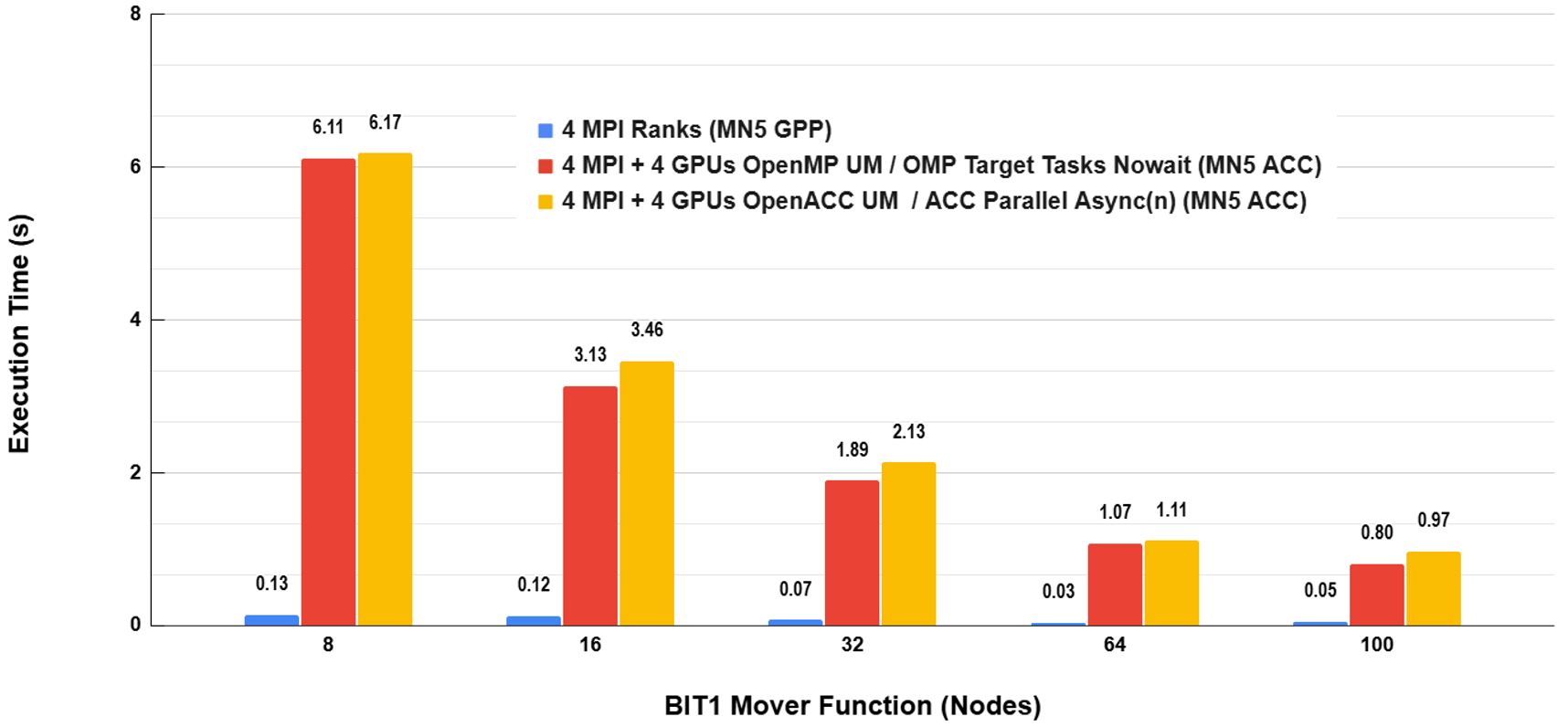}
        \caption{BIT1 Optimized mover function (strong scaling) execution time(s) for 200 time steps up to 100 Nodes (400 MPI Ranks and 400 GPUs) using OpenMP and OpenACC Unified Memory on \emph{MN5 H100} NVIDIA GPUs.} \label{BIT_GPU_Unified_Memory_Mover_Execution_Time_MN5_Extreme}
     \end{center}
\end{figure*}

For the total simulation and shown in Fig.~\ref{BIT_GPU_Unified_Memory_Total_Simulation_Execution_Time_MN5_Extreme}, OpenACC Async(n) achieves a runtime reduction from 23.24s on 8 nodes (32 GPUs) to 10.70s on 100 nodes (400 GPUs), while OpenMP Nowait improves from 23.11s to 10.43s over the same scaling range. The mover function and shown in Fig.~\ref{BIT_GPU_Unified_Memory_Mover_Execution_Time_MN5_Extreme}, a critical computational component, also shows performance gains across both implementations, with execution times decreasing from 6.17s to 0.97s for OpenACC and from 6.11s to 0.80s for OpenMP. However, as the node count increases, inter-node communication overhead and synchronization costs become more apparent, limiting further scaling benefits. 

On the CPU partition, execution times increase as the node count rises, showing diminishing returns in strong scaling due to communication overhead, memory contention, and I/O bottlenecks~\cite{williams2023leveraging}. For instance, CPU execution initially improves, with times decreasing from 3.96s on 8 nodes to 2.42s on 16 nodes but then rises to 8.74s on 100 nodes. A similar trend appears in the mover function. Despite the CPU being slightly faster than the GPUs at smaller scales, its efficiency declines at larger node counts, emphasizing the need to further optimize GPU implementations for extreme-scale performance. 

Evaluating speedup and parallel efficiency (PE) is essential when optimizing simulations on pre-exascale systems like MN5~\cite{banchelli2025introducing} and preparing for exascale systems like JUPITER~\cite{dumiak2023exascale}. While execution time shows overall performance, speedup measures the gain from added resources, and parallel efficiency evaluates resource utilization. These metrics are critical for identifying bottlenecks, ensuring scalability, and optimizing performance on advanced HPC architectures, which is crucial as we move toward exascale to tackle complex scientific and industrial challenges.

\begin{figure*}[h!]
    \begin{center}
        \includegraphics[width=\textwidth]{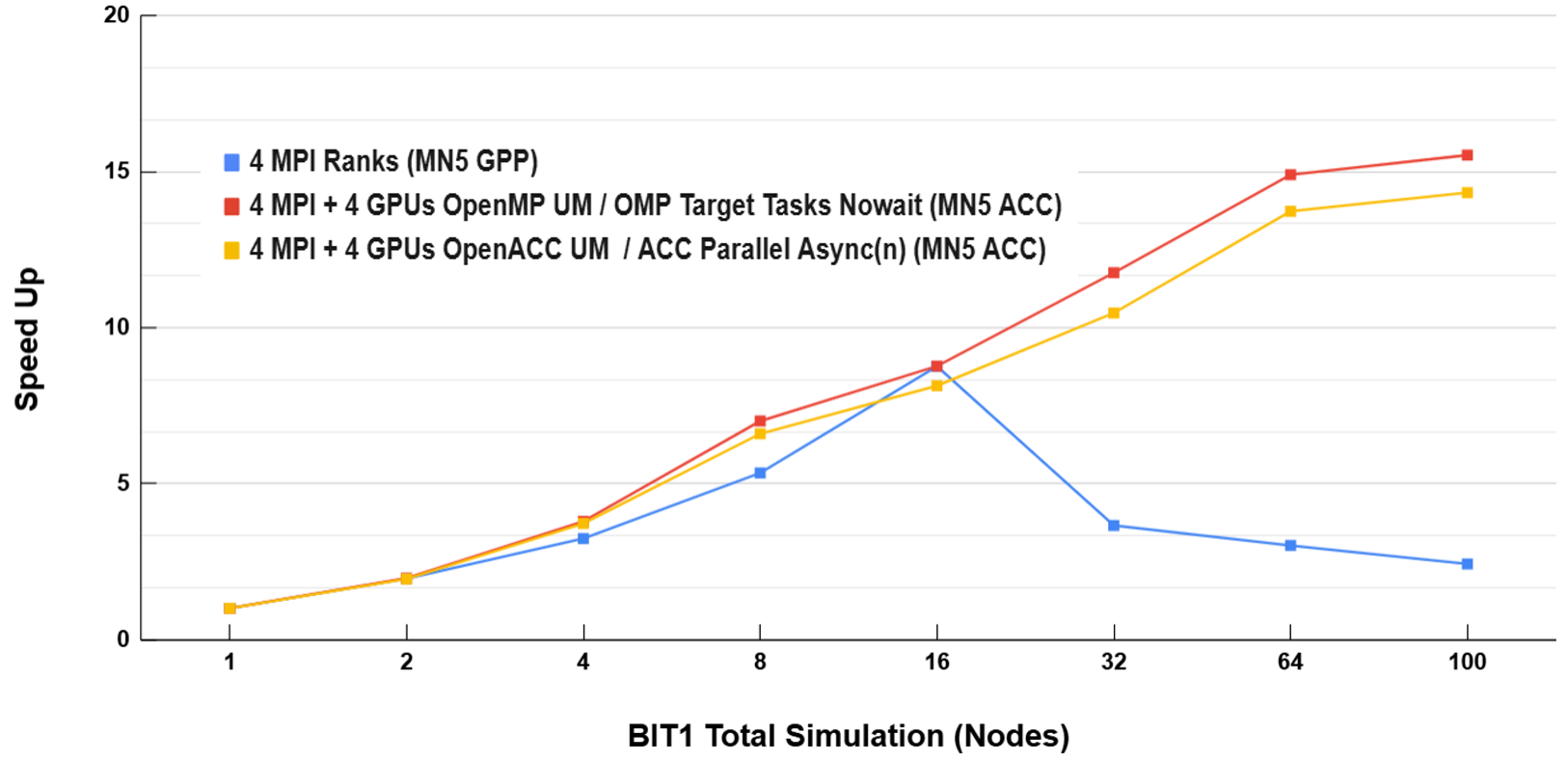}
        \caption{BIT1 Total Simulation Speed Up for 200 time steps scaling up to 100 Nodes (400 MPI Ranks and 400 GPUs) using OpenMP and OpenACC Unified Memory on \emph{MN5 H100} NVIDIA GPUs.} \label{MN5-BIT1-Speed-Up-Total-Simulations-400_GPUs_Scaling_Async}
     \end{center}
\end{figure*} 

\begin{figure*}[h!]
    \begin{center}
        \includegraphics[width=\textwidth]{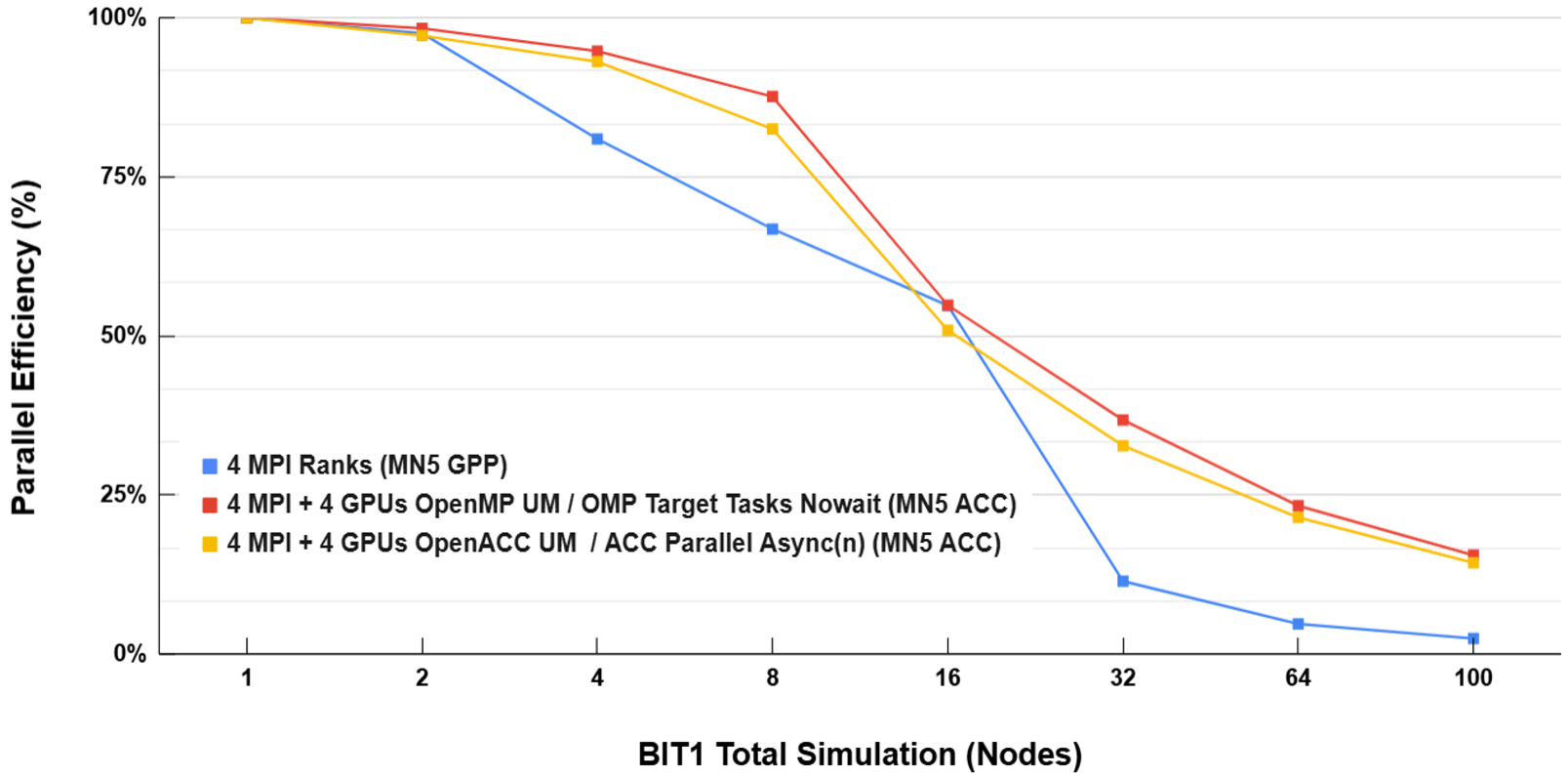}
        \caption{BIT1 Total Simulation Parallel Efficiency (\%) for 200 time steps scaling up to 100 Nodes (400 MPI Ranks and 400 GPUs) using OpenMP and OpenACC Unified Memory on \emph{MN5 H100} NVIDIA GPUs.} \label{MN5-BIT1-Parallel-Efficiency-Total-Simulations-400_GPUs_Scaling_Async}
     \end{center}
\end{figure*} 

In Fig.~\ref{MN5-BIT1-Speed-Up-Total-Simulations-400_GPUs_Scaling_Async} and Fig.~\ref{MN5-BIT1-Parallel-Efficiency-Total-Simulations-400_GPUs_Scaling_Async}, the CPU-only MPI version on MN5 GPP demonstrated solid scaling up to 16 nodes, achieving an 8.76$\times$ speedup (54.76\% PE), but suffered severe degradation beyond this point due to communication overhead, dropping to a 2.42$\times$ speedup (2.42\% PE) at 100 nodes. In contrast, asynchronous multi-GPU implementations on MN5 ACC exhibited significantly better scalability, with the OpenMP Nowait version reaching a 15.54$\times$ speedup (15.54\% PE) at 100 nodes, while the OpenACC Async(n) version achieved a 14.33$\times$ speedup (14.33\% PE). Up to 16 nodes, both asynchronous multi-GPU versions continue to show superior scalability, with OpenMP Nowait reaching 8.77$\times$ speedup (54.81\% PE) and OpenACC Async(n) attaining 8.14$\times$ speedup (50.87\% PE), surpassing the CPU-only MPI version in both speedup and parallel efficiency. At higher node counts, parallel efficiency declined across all implementations due to increased communication and synchronization costs. However, the asynchronous multi-GPU versions sustained significantly better parallel efficiency than the CPU-only MPI implementation, demonstrating the benefits of asynchronous multi-GPU execution in mitigating some scalability bottlenecks.

While the GPU-accelerated versions of BIT1, particularly the 4 MPI + 4 GPUs OpenACC UM / ACC Parallel Async(n) configuration, demonstrate performance improvements, the CPU version remains the fastest at smaller node counts. As node counts increase, OpenMP configurations, such as the 4 MPI + 4 GPUs OpenMP UM / OMP Target Tasks Nowait configuration, exhibit better scalability and performance, making OpenMP the preferred option for large-scale PIC simulations where high throughput and efficient GPU utilization across multiple nodes are critical. However, despite these promising results, a critical challenge arises from data transfer constraints during each PIC cycle, as seen in Fig.~\ref{diagram}. Profiling results, illustrated in Fig.~\ref{BIT1_Unified_Memory},~\ref{BIT1_Profiling_Vega}, and~\ref{BIT1_Async_Profiling_MN5}, highlight performance bottlenecks caused by substantial data transfers from CPU to GPU at each timestep. To address this, minimizing large data transfers between the CPU and GPU, as well as modifying the original data layout~\cite{tskhakaya2007optimization}, is essential. Exploring CUDA streams and utilizing particle batch processing with OpenMP configurations across multiple GPUs per node presents a promising strategy to streamline both data transfer and computation, as noted in~\cite{chien2020sputnipic}.


\section{Related Work}
BIT1, an advanced PIC code, is designed to simulate plasma-material interaction~\cite{tskhakaya2010pic}, and has been used in
various applications, including fusion devices such as tokamaks. It builds upon the XPDP1 code~\cite{verboncoeur1993simultaneous}, initially developed by Verboncoeur's team at Berkeley, incorporating optimized data layout to efficiently handle collisions~\cite{tskhakaya2007optimization}. Recently, Williams \etal~\cite{williams2023leveraging} highlighted the particle mover as one of the most computationally intensive parts of BIT1. Several studies have
explored hybrid parallelization techniques, using MPI and OpenMP for PIC codes, including Smilei~\cite{derouillat2018smilei},
iPIC3D~\cite{markidis2016epigram,peng2015acceleration,williams2024characterizing}, and Warp-X~\cite{vay2018warp}. In contrast to earlier approaches, task-based shared-memory parallelization~\cite{ayguade2008design,liu2023parallel} has been employed to specifically address load-imbalance issues in BIT1's particle mover~\cite{williams2023leveraging}.
Expanding on this work, Williams \etal~\cite{williams2024optimizing} implemented a hybrid shared-memory version of BIT1, utilizing OpenMP and OpenACC~\cite{chandrasekaran2017openacc,markidis2015openacc,rossi2012towards,wei2016performance} to enable GPU acceleration. Their implementation introduced the first GPU-capable version of BIT1, focusing on data movement strategies like unified memory and explicit data transfer, achieving further optimization through concurrent GPU utilization. Other research has explored asynchronous execution models to enhance GPU offloading, such as Guaitero \etal~\cite{guaitero2022automatic}, who introduced automatic asynchronous execution of OpenMP offloaded regions, reducing complexity and improving performance. Kelling \etal~\cite{kelling2021challenges} addressed portability challenges in large-scale scientific applications by porting the
GPU-accelerated PIConGPU to OpenACC and OpenMP target, adding new backends to the Alpaka offloading abstraction for code portability. Valero-Lara \etal~\cite{valero2021openmp} introduced OpenMP Target Task, integrating tasking and GPU offloading into a unified model, demonstrating significant speedups in heterogeneous systems. Tramm \etal~\cite{tramm2022toward} ported the CPU-based OpenMC~\cite{romano2015openmc} Monte Carlo code to GPUs using the OpenMP target offload model, achieving performance equivalent to 200 Xeon CPU cores on an NVIDIA A100 for a reactor benchmark. Matsumura \etal~\cite{matsumura2021jacc} presented JACC, an OpenACC runtime framework that supports kernel-level and multi-GPU parallelization, achieving nearly linear scaling on multi-GPU systems. Markidis \etal~\cite{markidis2016epigram} ported NekBone to GPUs with OpenACC, achieving 43 Gflops on a single node and 79.9\% parallel efficiency on 1024 GPUs. Additionally, Xu \etal~\cite{xu2015multi} addressed the challenges of programming on heterogeneous systems, proposing a task-based extension of OpenACC for multi-GPU support, demonstrating significant performance improvements through case studies on scientific applications. Xu \etal~\cite{xu2024particle} used OpenACC for particle tracking in thermal flows, improving parallel efficiency from 39.8\% to 77.5\% on 8 GPUs through memory optimizations, maximizing GPU resource utilization and demonstrating OpenACC's suitability for PIC workloads by balancing fine-grained parallelism with GPU constraints.

\section{Discussion and Conclusion}

Asynchronous multi-GPU programming plays a critical role in optimizing the performance of the BIT1 code, particularly in enhancing the particle mover function. By taking advantage of overlapping computation and communication, we can achieve more efficient utilization of available GPU resources, reducing idle time and improving overall execution efficiency. The implementation of OpenMP Target Tasks with the "nowait" and "depend" clauses, alongside OpenACC parallel regions using the "async(n)" clause, allows for concurrent execution across multiple GPUs, maximizing throughput and leveraging the full computational power of the hardware.

Our primary goal was to enhance the particle mover function in BIT1, focusing on node-level efficiency and GPU offloading. Hybrid MPI and OpenMP/OpenACC approaches significantly improved on-node performance, showcasing the potential to efficiently leverage multicore CPUs and GPUs.


\begin{lstlisting}[
    language=C,
    style=mystyle,
    basicstyle=\fontsize{6}{6}\selectfont\ttfamily,
    caption={Simplified C code snippets illustrating the data structure and memory layout from the initial investigation of moveb(), one of the most computationally intensive functions in BIT1.},
    captionpos=b,
    label=BIT1_Data_Layout_Code_Moveb,
    float=!ht,
    tabsize=2,
    xleftmargin=2em, 
    framexleftmargin=2em
]
void moveb() {

    // Particle properties stored in multi-dimensional arrays (e.g., x[isp][j][i], vx[isp][j][i])
    
    atemp = a[j] + x[isp][j][i] * (a[j+1] - a[j]);  // Compute based on position
    vx[isp][j][i] += atemp;                         // Update velocity
    ...
    ...
    x[isp][j][i] += nstep[isp] * vx[isp][j][i];     // Update position
    
}

void moveb_VoS() {

    // Particles stored in a vector of particle objects (e.g., p.x, p.vx)
    
    atemp = a[j] + p.x * (a[j+1] - a[j]);           // Compute based on position
    p.vx += atemp;                                  // Update velocity
    ...
    ...
    p.x += nstep[sp] * p.vx;                        // Update position
    
}

void moveb_AoS() {

    // Particles stored in an array of particle structures (e.g., part[i].x, part[i].vx)
    
    atemp = a[j] + part[i].x * (a[j+1] - a[j]);     // Compute based on position
    part[i].vx += atemp;                            // Update velocity
    ...
    ...
    part[i].x += nstep[isp] * part[i].vx;           // Update position
    
}
\end{lstlisting} 

Results from multicore CPUs demonstrated the effectiveness of the hybrid approaches. The scalability of the MPI+OpenMP version indicates its potential for large-scale plasma simulations, which are crucial for efficient use of current supercomputer infrastructure.

GPU porting using OpenMP and OpenACC unveiled challenges and opportunities in tapping into GPU resources for the first time. Emphasizing a balanced hybrid approach for optimal GPU performance, our findings suggest that implementing OpenMP or OpenACC on GPUs may increase execution times, potentially outweighing parallel processing benefits. Notably, the OpenMP Target with 2 GPUs demonstrated a significant reduction in execution time among GPU results, highlighting potential performance improvement through concurrent GPU utilization, especially when dedicated GPUs are assigned to MPI ranks.

Initially, the fastest GPU execution was achieved with the 4 MPI + 4 GPUs OpenACC UM / ACC Parallel Async(n) configuration, highlighting the importance of asynchronous multi-GPU programming in improving performance, especially for short runs. However, for extreme-scale runs, OpenMP configurations, such as the 4 MPI + 4 GPUs OpenMP UM / OMP Target Tasks Nowait, offered better scalability and performance, making OpenMP the preferred option for large-scale PIC simulations.

\begin{figure*}[h!]
    \begin{center}
        \includegraphics[width=\textwidth]{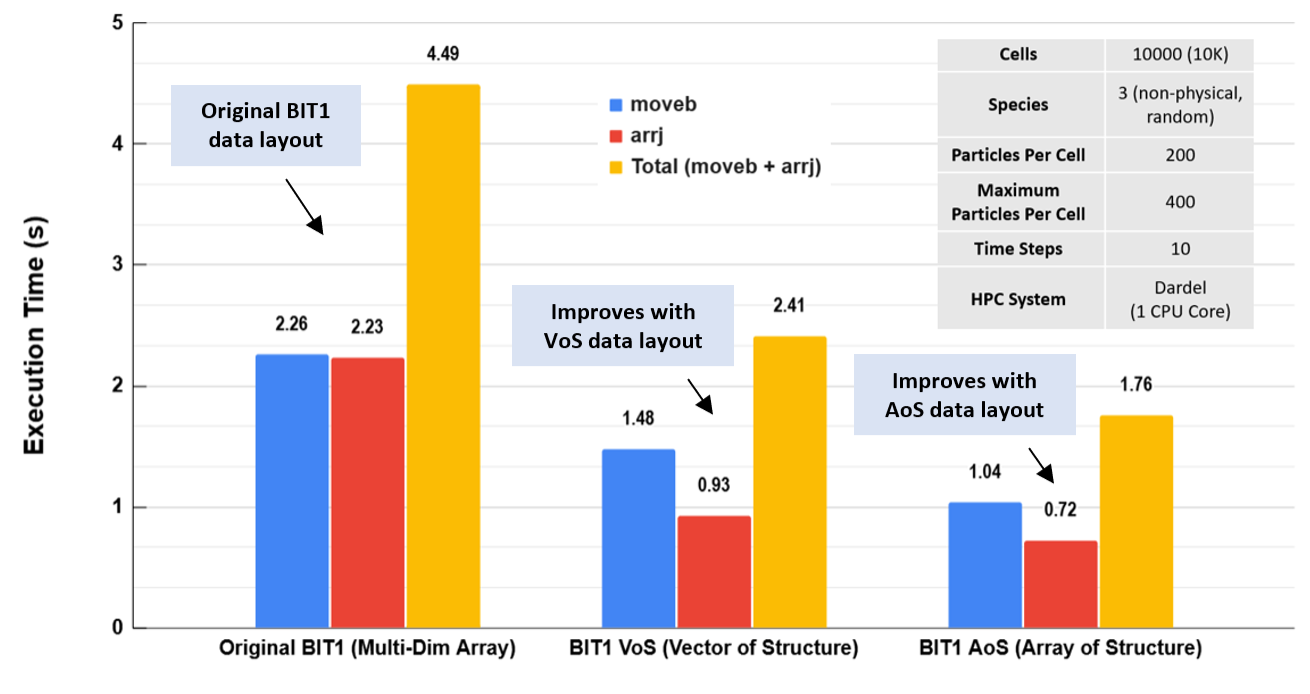}
        \caption{An initial investigation of the data structure and memory layout for two of the most computationally intensive BIT1 functions on \emph{Dardel}.} \label{BIT1_Data_Layout_Study}
     \end{center}
\end{figure*}

Future research to further enhance GPU performance in BIT1, as shown in Listing~\ref{BIT1_Data_Layout_Code_Moveb} and Fig.~\ref{BIT1_Data_Layout_Study}, will focus on optimizing the data structure and memory layout. For most BIT1 production cases, \texttt{moveb()} is one of the most frequently used functions~\cite{williams2023leveraging}. As mentioned earlier and shown in Fig.~\ref{BIT1_sorting}, the original \texttt{moveb()} function uses arrays to store particle properties, requiring more memory and leading to inefficient memory access patterns when offloaded to the GPU. Transitioning to a vector of particle objects, \texttt{moveb\_VoS()}, offers dynamic resizing and easier access to particle attributes on the GPU~\cite{kolb2005dynamic,meneses2022ggarray,tang2014accelerating}. Incorporating dynamic memory allocation into BIT1 further enhances scalability and flexibility by enabling more efficient memory management based on the number of particles and system resources, optimizing GPU performance, and adapting better to various hardware configurations. Additionally, using an array of particle structures, \texttt{moveb\_AoS()}, ensures contiguous memory access, improving memory throughput and reducing latency, which better aligns with GPU processing~\cite{homann2018soax,strzodka2012abstraction}.

Adopting the Low-Level Abstraction for Memory Access (LLAMA) framework can further enhance BIT1's memory access efficiency. LLAMA, a C++ library, facilitates efficient handling of multidimensional arrays and custom memory layouts at compile time~\cite{gruber2023llama}. Integrating LLAMA allows BIT1 to optimize memory layouts for specific hardware, leading to improved memory access patterns and accelerated GPU performance.

Enhanced hybrid MPI, OpenMP, and OpenACC approaches provide comprehensive parallelization, alleviating I/O bottlenecks. Investigating the use of openPMD with Hybrid BIT1 and GPU offloading will further address the I/O bottlenecks previously identified in~\cite{williams2023leveraging}, fixed in~\cite{williams2024parallelio}, and reviewed in~\cite{williams2024understanding}. This further integration will improve large dataset handling, enhance data management efficiency, and support GPU workloads in BIT1, enabling it to adapt to a wide range of computing environments. 

\vspace{2mm} 
\noindent \small{\textbf{Acknowledgments.} Funded by the European Union. This work has received funding from the European High Performance Computing Joint Undertaking (JU) and Sweden, Finland, Germany, Greece, France, Slovenia, Spain, and Czech Republic under grant agreement No 101093261 (Plasma-PEPSC). The computations/data handling were/was enabled by resources provided by the National Academic Infrastructure for Supercomputing in Sweden (NAISS), partially funded by the Swedish Research Council through grant agreement no. 2022-06725.}





\bibliographystyle{splncs04}

\end{document}